\documentclass[10pt,journal]{IEEEtran}
\usepackage{amsfonts}
\usepackage{amssymb}
\usepackage{sublabel}

\usepackage[noadjust]{cite}

\ifCLASSINFOpdf
  \usepackage[pdftex]{graphicx}
\else
  \usepackage[dvips]{graphicx}
\fi
\usepackage[cmex10]{amsmath}
\usepackage[tight,footnotesize]{subfigure}

\usepackage{caption}

\begin{document}
%
\title{Easily Computed Lower Bounds on the Information Rate
of Intersymbol Interference Channels
}





%

\author{
Seongwook Jeong, \IEEEmembership{Student Member, IEEE,}
and Jaekyun Moon, \IEEEmembership{Fellow, IEEE}
\IEEEcompsocitemizethanks{
\IEEEcompsocthanksitem{This work was supported in part by the NSF under Theoretical Foundation grant no. 0728676
and the National Research Foundation of Korea under grant no. 2010-0029205.}
\IEEEcompsocthanksitem {S. Jeong is with the Dept. of Electrical and Computer Engineering,
University of Minnesota, Minneapolis, MN 55455 USA (e-mail: jeong030@umn.edu)}.
\IEEEcompsocthanksitem {J. Moon is with Dept. of Electrical Engineering, Korea Advanced Institute of Science and Technology, Daejeon, 305-701, Republic of Korea (e-mail: jmoon@kaist.edu).}%
}
}

\markboth{TO APPEAR IN IEEE INFORMATION THEORY}{S. Jeong and J. Moon: Easily Computed Lower Bounds on the Information Rate
of Intersymbol Interference Channels}


\maketitle
\setlength\arraycolsep{1pt}
\thispagestyle{empty}

\begin{abstract}
Provable lower bounds are presented for the information rate $I(X;
X+S+N)$ where $X$ is the symbol drawn independently and uniformly
from a finite-size alphabet, $S$ is a discrete-valued random
variable (RV) and $N$ is a Gaussian RV. It is well known that with $S$ representing the
precursor intersymbol interference (ISI) at the decision feedback equalizer (DFE) output,
$I(X; X+S+N)$ serves as a tight lower bound for the symmetric information rate (SIR) as well as
capacity of the ISI channel corrupted by Gaussian noise.
When evaluated on a number of well-known finite-ISI channels, 
these new bounds provide a very similar level of tightness against the SIR 
to the conjectured lower bound by Shamai and Laroia at all signal-to-noise ratio (SNR) ranges,
while being actually tighter when viewed closed up at high SNRs.
The new lower bounds are obtained in two steps: First, a ``mismatched"
mutual information function is introduced which can be proved as a lower bound to $I(X; X+S+N)$.
Secondly, this function is further bounded from below by an expression that can be computed easily via a few single-dimensional
integrations with a small computational load. 
\end{abstract}

\begin{keywords}
Channel capacity, decision feedback equalizer, information rate, intersymbol interference, lower bounds, mutual information.
\end{keywords}

%
\setcounter{page}{1}
\newtheorem{definition}{Definition}
\newtheorem{proposition}{Proposition}
\newtheorem{lemma}{Lemma}

\section{Introduction}
The computation of the symmetric information rate (SIR) of the
classical discrete-time intersymbol interference (ISI) channel is
of great interest in digital communication. The SIR represents the
mutual information between the channel input and output while the
input is constrained to be independently and uniformly distributed
(i.u.d.) over the given alphabet. In this sense, the SIR is also
known as capacity with uniform, independent input distribution and
itself represents a reasonably tight lower bound to unconstrained channel
capacity, especially at high coding rates. During recent years, a number of researchers have worked
on estimating or bounding the information rate via simulation of
the Bahl-Cocke-Jelinek-Raviv (BCJR) algorithm \cite{BCJR74}. The
information rate with a given input distribution can be closely
estimated for finite ISI channels with moderate input alphabet
size and channel impulse response length, by running the
forward-recursion portion of the BCJR algorithm on long (pseudo)
randomly generated input and noise samples \cite{Arnold01},
\cite{Sharma01}, \cite{Pfister01}. The simulation-based method has
been further generalized, and lower and upper bounds based on
auxiliary finite-state channels with reduced states were introduced for
long ISI channels, as well as some non-finite state ISI channels
in \cite{Arnold06}. The tightness of these bounds is highly
related to the optimality of auxiliary channels, but the general
rule to find the optimal or near-optimal auxiliary channel has not
been provided in \cite{Arnold06}. The work of \cite{Arnold06} has
been recently extended in \cite{Sadeghi09} to further tighten the
lower and upper bounds by using an iterative
expectation-maximization type algorithm to optimize the parameters
of the auxiliary finite-state channels. It is noted, however, that the
global optimality of the bounds in \cite{Sadeghi09} is neither
guaranteed, nor the lower bound is proven to converge to a
stationary point as iteration progresses. Another approach based
on auxiliary channels is also proposed to obtain a lower bound
utilizing a mismatched Ungerboeck-type channel response to achieve
improved tightness for a given level of computational complexity
\cite{Rusek09}. In the context of \cite{Rusek09}, the
Ungerboeck-type response is the channel's response observed at the
output of the matched filter front-end. As such, the trellis
search detection algorithms driven by the channel observations of
the Ungerboeck model must be designed so that they can handle
correlated noise samples \cite{Ungerboeck}.

An entirely different direction in estimating or bounding the
information rate is based on finding an analytical expression that
can easily be evaluated or numerically computed (in contrast to
the methods based on Monte-Carlo simulation that rely on
generating pseudo-random signal and noise samples). An early work
in this direction is the lower bound on the SIR by Hirt
\cite{Hirt88} based on carving a fixed block out of the channel
input/output sequences and performing a single multi-dimensional
integration (or running Monte-Carlo simulation for estimating the
integral) with the dimensionality equal to the block size.
However, this method is also computationally intense unless the
size of the block gets small. Unfortunately the lower bound of
\cite{Hirt88} is not tight unless the block size is very large
compared to the channel ISI length.

A number of more computationally efficient and analytically
evaluated lower bounds for the SIR have been discussed in
\cite{Shamai91}, \cite{Shamai96}. Unfortunately, however, the only
bound presented in \cite{Shamai96} that is reasonably tight
throughout the entire signal-to-noise ratio (SNR) region (i.e.,
both low and high code rate regimes) is the one that could not be
proved as a lower bound. This particular bound is now widely known
as the Shamai-Laroia conjecture (SLC) and, although unproven, is a
popular tool for quickly estimating the SIR of ISI channels. At
high code rates, the SIR is generally very close to capacity, so
an easily computed tight SIR lower bound is also useful for quickly
estimating channel capacity for high code rate applications, such
as data storage channels and optical fiber channels. 

Consider the
random variable (RV) $Y = X + S + N$, where $X$ is a symbol drawn
independently and uniformly from a fixed, finite-size alphabet set
symmetrically positioned around the origin, $S$ a zero-mean
discrete-valued RV, and $N$ a zero-mean Gaussian RV. The SLC is
concerned with the special case where $S$ is a linear sum of
symbols drawn independently and uniformly from the same symbol set
where $X$ was taken. As the number of symbols forming $S$ grows,
finding an analytical expression for the probability density
function of $S + N$ (and thus one for $I(X; Y)$) is a
long-standing problem \cite{Garsia63}, \cite{Wittke88}, as pointed
out in \cite{Shamai96}. The SLC of \cite{Shamai96} can be stated
as $I(X; X+S+N) \geq I(X; X+G)$, where $G$ is a Gaussian RV with
variance matching that of $S+N$.  
The information rate $I(X; X+G)$
is easily obtained by numerically calculating a single
one-dimensional integral, and is generally observed to be
reasonably tight to $I(X; X+S+N)$ in most cases. Unfortunately,
$I(X; X+G)$ remains as a conjectured bound with no proof available
to date. One difficulty of proving the SLC stems from the fact
that for the channels driven by the inputs from a finite alphabet,
Gaussian noise is not the worst-case noise in terms of the
achievable information rate \cite{Shamai96},
\cite{Shamai-Verdu92}. Another difficulty is that the power
contribution of a single individual weight involved in
constructing $S$ could remain a significant portion of the total
power associated with all weights, even if the number of weights
approaches infinity. This is to say that the Lindberg condition
for the central limit theorem does not hold for this problem, and
the Gaussian approximation of $S$ cannot be justified
\cite{Shamai96}.

In this paper, we are also interested in the easily computable
analytical expressions for lower bounds to $I(X; X+S+N)$.
Note that, in the context of the unbiased minimum mean-squared-error decision
feedback equalizer (MMSE-DFE) application, $S$ represents
the collection of residual precursor ISI contributions
and in this case $I(X; X+S+N)$ itself is a well-known lower bound to 
the SIR \cite{Shamai96}. The
bounds we develop here are fairly tight, with their tightness
generally enhanced with increasing computational load (which in
the end still remains small). Our approach is to first define a
``mismatched" mutual information (MI) function based on the
``mismatched" entropy that takes the $\log$ operation not on the
actual underlying probability density but on the Gaussian density
with the same variance. We then prove that this ``mismatched" MI
is always less than or equal to $I(X; Y)$. We further
bound this function from below so that the final bound can be
evaluated using numerical integration. The bound is basically
evaluated by computing a few single-dimensional integrals. This is
in contrast to the Hirt bound that computes a single
multi-dimensional integral of very high dimension. Our bound
computation also requires the evaluation of sum of the absolute
values of the linear coefficients that form $S$ as well as the
identification of dominant coefficient values, if they exist. 
With the application of the MMSE-DFE, these linear coefficients correspond to the
weights on the interfering symbols after ideal postcursor ISI
cancellation and can easily be obtained with a small amount of computation.
At a reasonable overall
computational load, our bounds are shown to be for all practical purposes as tight
as the Shamai-Laroia conjecture for many practical ISI channels.

Section \ref{sec:MMI_SIR} presents the provable bound to $I(X;Y)$ and numerically
compares it with the SLC for some example distributions for the
linear coefficients that form $S$. Section \ref{sec:LB_MMI} develops upper and
lower bounds on the provable bound itself, based on identifying
clusters in the distribution of $S+N$. Finding clusters in the
$S+N$ distribution is the same as identifying dominant coefficient
values from the linear coefficient set that is used to construct
$S$. Section \ref{sec:Numerical Results} generates and discusses numerical results. In all
finite-ISI channels examined, our bound provides the same level of
tightness as the SLC against the SIR (while being actually tighter than
SLC at high SNRs when viewed closed up) with a very reasonable
computation load. In particular, our lower bound is presented on
the same channel employed in \cite{Sadeghi09}. This provides an
indirect means to compare the computational loads of our method and that of 
\cite{Sadeghi09}. As expected, our analytical method is considerably better in
quickly producing a reasonably tight bound than the simulation-based method
of \cite{Sadeghi09} in terms of complexity/accuracy tradeoffs.
Note that the method of \cite{Sadeghi09} represents the latest
development in simulation-based SIR bounds. Section \ref{sec:Conclusion} concludes
the paper.

\section{A Provable Lower Bound to the Symmetrical Information Rate}\label{sec:MMI_SIR}
We first present a provable lower bound to $I(X;Y)$ where $Y = X +
\sum_{k=1}^{L} d_{-k}X_{k} + N = X + S + N$. The symbols $X$ and
$X_k$ are all independently and uniformly drawn. The linear
coefficients $d_{-k}$'s are related to the channel impulse
response and will be specified in Section \ref{sec:Numerical Results}. Let $V=S+N$ so we
can write $Y=X+V$. Note that $V$ is a Gaussian mixture. Also let
$Z=X+G$ where $G$ is a zero mean Gaussian with variance matching
that of $V$, i.e., $\sigma_G^2 = \sigma_V^2$.

\begin{definition}[``Mismatched" MI (MMI) Function] 
Define
\begin{eqnarray}
I'(X; Y) & \triangleq & H'(Y) - H'(V)
\end{eqnarray}
where
\begin{eqnarray}
H'(Y)  & \triangleq & - \int_{-\infty}^{\infty} f_Y(t) \log f_{Z}(t) dt , \nonumber \\
H'(V)  & \triangleq & - \int_{-\infty}^{\infty} f_V(t) \log f_{G}(t) dt \nonumber
\end{eqnarray}
and $f_Y(t)$, $f_V(t)$, $f_Z(t)$, and $f_G(t)$ are the probability
density functions (pdfs) of the RVs, $Y$, $V$, $Z$, and $G$,
respectively. Note that the ``mismatched" entropy functions
$H'(Y)$ and $H'(V)$ are defined based the $\log$ operation applied not to
the actual underlying pdf $f_V(t)$ but rather to the ``mismatched"
Gaussian pdf $f_G(t)$.

\end{definition}

\begin{lemma}\label{lemma1}
Given the MMI function defined as above, we have
\begin{eqnarray}
I'(X; Y) \leq I(X; Y).
\end{eqnarray}

\begin{proof}
See Appendix \ref{proof:lemma1}.
\end{proof}
\end{lemma}

Let us now take a close look at this MMI function $I'(X; Y)$ and
develop some insights into its behavior. Let the variances of
$V$, $S$, and $N$ be $\sigma_V^2$, $\sigma_S^2$, and $\sigma_N^2$
respectively. Further assume that the RVs, $X$, $V$, $S$, and $N$
are all real-valued. We will also assume a binary input alphabet.
These assumptions are not necessary for our development but make
the presentation clearer as well as less cluttered. We will simply
state the results in Section \ref{subsec:LB_Complex} for a non-binary/complex-valued
example. We also denote $m_i = \sum_{k=1}^{L} d_{-k}X_{k}$ for
$i=1,2,\ldots, 2^L$ since $\{X_{k}\}_{k=1}^{L}$ can have $2^L$
different sequences. Naturally, the pdfs of RVs $V$ and $G$ can be
written as
\begin{eqnarray}
f_V(t) & = & 2^{-L} \sum_{i =1}^{2^L} \dfrac{1}{\sqrt{2 \pi \sigma_{N}^2} } \exp \left( - \dfrac{ \left( t - m_i \right )^2 } {2 \sigma_{N}^2 } \right)  \nonumber \\
f_G(t) & = &  \dfrac{1}{ \sqrt{2 \pi  \sigma_{V}^2} } \exp \left( - \dfrac{ t^2 } { 2 \sigma_{V}^2 } \right). \nonumber
\end{eqnarray}

\begin{proposition}
Denoting $\rho_i \triangleq m_i /\sqrt{P_X}$ and $\tau \triangleq (t-m_i)/\sigma_N$, letting $\rho_k^+$'s
to mean the positive-half subset of $\rho_i$'s, and defining $R
\triangleq P_X / \sigma_V^2$ and $\phi \triangleq \sigma_N /
\sigma_V$, the MMI function can be rewritten as $I'(X; Y) = \log 2
-F$ with the new definition 
\sublabon{equation}\begin{eqnarray}
F & \triangleq & 2^{-L} \sum_{i=1}^{2^L} \mathrm{E}_{\tau} \left[ \log \left \{ 1 + e^{-2 R \rho_i} e^{-2 \phi \sqrt{R} \tau - 2R}  \right \} \right] \nonumber \\
& = & \mathrm{E}_{\rho, \tau} \left[ \log \left \{ 1 + e^{-2 R \rho} e^{-2 \phi \sqrt{R} \tau - 2R}  \right \} \right]  \label{eq:F1} \\
& = & 2^{ - (L - 1)} \sum_{k = 1}^{2^{L - 1} } \mathrm{E}_{\tau} \bigg[ \frac{1}{2} \log \Big\{ 1 + 2\cosh \left( {2R\rho_k^+  } \right) e^{ - 2\phi \sqrt{R} \tau  - 2R} \nonumber \\
& & \quad  + e^{ - 4\phi \sqrt{R} \tau  - 4R}  \Big\}  \bigg] \nonumber \\
& = & \mathrm{E}_{\rho^+, \tau} \bigg[ \frac{1}{2}\log \Big\{ 1 + 2 \cosh \left( {2R\rho^+  } \right) e^{ - 2\phi \sqrt{R} \tau  - 2R}  \nonumber \\
& & \quad + e^{ - 4\phi \sqrt{R} \tau  - 4R}  \Big\} \bigg]. \label{eq:F2}
\end{eqnarray}\label{proposition1}\sublaboff{equation}
\end{proposition}

A detailed derivation is given in Appendix
\ref{appendix:proposition1}. The position $m_i$ of the $i$th
Gaussian pdf of the mixture $f_V(t)$ is expressed as a
dimensionless quantity: $\rho_i = m_i/\sqrt{P_X}$, with the
normalization by the square root of the input power. Because of
the symmetric nature of $f_V(t)$, $\rho_i$ occurs in
equal-magnitude, opposite-polarity pairs. The expectation is
initially over $\tau$, which is considered a
zero-mean unit-variance Gaussian random variable when contained
inside the argument of the expectation operator. The expectation
operator in this case can simply be viewed as a short-hand
notation as in
\begin{eqnarray}
\mathrm{E}_{\tau} \left[ p(\tau) \right] = \int_{-\infty}^{\infty} \dfrac{e^{-\tau^2 / 2}}{\sqrt{ 2 \pi } }  p(\tau) d\tau . \nonumber
\end{eqnarray}
In (\ref{eq:F1}) and (\ref{eq:F2}), however, $\rho$ (or $\rho^+$)
is also treated as a RV and the expectation is over both $\tau$
and $\rho$ (or $\tau$ and $\rho^+$) as the double subscripts
indicate. Given the pdfs of $\tau$, $\rho$ and $\rho^+$, the
computation of the expectation now involves numerical evaluation
of a double integral. Note that in (\ref{eq:F1}) $\rho$ is a
discrete-valued random variable distributed according to
$f_{\rho}(t)$, which denotes the probability distribution of $\rho
= (1 / \sqrt{P_X} ) \sum_{k=1}^{L} d_{-k}X_k$ and $\rho^+$ is a
discrete-valued random variable distributed according to $2
f_{\rho}(t) u(t)$ where $u(t)$ is a step function. 
Also, notice that
$\cosh ( {2R\rho^+  } ) \geq 1$ and $\phi \leq 1$.
Since it is not easy to find $f_{\rho}(t)$ when $L$ is large, 
evaluating (\ref{eq:F1}) or (\ref{eq:F2}) is difficult in general.

It is insightful to compare $F$ with
\sublabon{equation}\begin{eqnarray}
F_{SLC} & \triangleq & \log 2 - C_{SLC} (R) \nonumber \\
&  = & \int_{ - \infty }^{\infty}  \dfrac{e^{ - \tau^2 /2}} { \sqrt {2\pi } }  \log \left \{ {1 + e^{ - 2\sqrt{R} \tau  - 2R} } \right\} d\tau  \nonumber \\
& = & \mathrm{E}_{\tau} \left[ \log \left\{ {1 + e^{ - 2\sqrt{R}\tau  - 2R} } \right\} \right] \label{eq:F_b1} \\
& = & \mathrm{E}_{\tau} \left[ \dfrac{1}{2}\log \left\{ 1 + 2e^{ - 2\sqrt{R} \tau  - 2R}  + e^{ - 4\sqrt{R}\tau  - 4R}  \right\} \right] \nonumber \\ \label{eq:F_b2}
\end{eqnarray}\sublaboff{equation}where $C_{SLC}(R)$ is the SIR of the binary-input Gaussian channel
with SNR given by $R \triangleq P_X / \sigma_V^2$ and is the
well-known SLC. The function $F_{SLC}$ quantifies the gap between
the SLC and the maximum attainable capacity for any binary channel
with infinite SNR, namely, 1 bit/channel use. Comparing
the expressions for $F$ in (\ref{eq:F2}) and $F_{SLC}$ in
(\ref{eq:F_b2}), we see that if $\rho^+ = 0$ so that $\phi =1$,
then $F = F_{SLC}$, and $I'(X; Y)$ and the SLC both become equal
to $I(X;Y)$. Also, if the discrete RV $\rho$ converges to a
Gaussian random variable (in cumulative distribution), then again
we get $F = F_{SLC}$ and $I'(X; Y)=C_{SLC}(R)=I(X;Y)$.

Furthermore, that $\rho^+ \geq 0$ in (\ref{eq:F2}) makes $F$
larger while the factor $\phi$ being less than 1 has an effect of
decreasing $F$ as it increases. If $I'(X;Y) = \log 2 - F$ is to be
a tight lower bound to $I(X; Y)$, then $F$ needs to be small. The
important question is: how does $F$ overall compare with
$F_{SLC}$, over all interested range of SNR? Since it is already
proved that $I'(X; Y) = \log 2 - F$, if $F \leq F_{SLC}$ for some
$R$ values, then clearly $C_{SLC}(R) = \log 2 - F_{SLC} \leq I(X;
Y)$ at those SNRs, i.e., the SLC holds true at least at these
SNRs. 

While exact computation of (\ref{eq:F2}) requires in general
obtaining all possible positive-side values of $\rho =
(1/\sqrt{P_X}) \sum_{k=1}^{L} d_{-k} X_k$ and thus can be
computationally intense for large $L$, in the cases where we know
the functional form of the distribution for $\rho$, evaluation of
(\ref{eq:F1}) or (\ref{eq:F2}) is easy; the behavior of $F$ under
different $\rho$ distributions offers useful insights.

\begin{figure}[!t]
\centering 
\includegraphics[width=8.5cm]{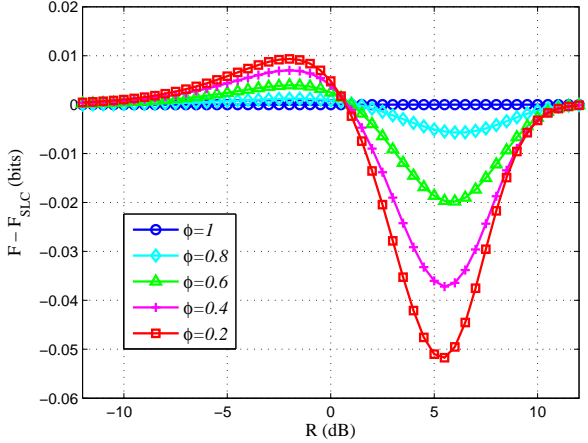}
\caption{$F-F_{SLC}$ as a function of $R$ for a uniform $\rho$.}
\label{fig:F_unif}
\end{figure}

\begin{figure}[!t]
\centering 
\includegraphics[width=8.5cm]{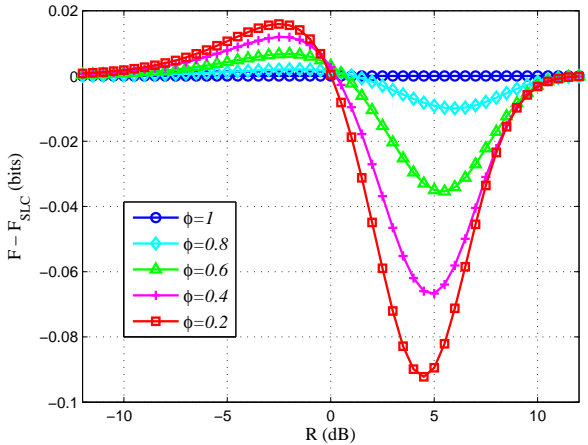}
\caption{$F-F_{SLC}$ as a function of $R$ for a two-valued $\rho$.}
\label{fig:F_onetap}
\end{figure}

First try a uniform distribution for $\rho$. For a uniformly
distributed discrete random variable $\rho$ from $-K\Delta = -
|\rho|_{\max}$ to $K\Delta =  |\rho|_{\max}$ with a gap $\Delta$
between delta functions in the pdf, we have
\begin{eqnarray}
 \sigma_S^2  & = &  \dfrac{2 P_X \Delta^2}{{2K + 1}} \sum_{i = 1}^K { i^2  }  = \dfrac{ {P_X \Delta^2 K(K + 1)} }{3} \nonumber \\
 & = & \dfrac{{P_X |\rho|_{\max } (|\rho| _{\max }  + \Delta )}}{3} \nonumber
\end{eqnarray}
which makes
\begin{eqnarray}
\phi^2  & = & \dfrac{{\sigma_N^2 }}{{\sigma_N^2  + \sigma_S^2 }} = 1 - \frac{{\sigma_S^2 }}{{\sigma_V^2 }} = 1 - \dfrac{R\Delta^2 K(K + 1)}{3} \nonumber \\
& = & 1 - \dfrac{R|\rho|_{\max } (|\rho|_{\max }  + \Delta )}{3}. \nonumber
\end{eqnarray}
Fig. \ref{fig:F_unif} shows $F$ and $F_{SLC}$ plotted with $K = 1000$
as functions of $R$ for various values of $\phi$. We also consider a simple case involving only a single coefficient
$d_{-1}$, in which case $\rho$ takes only two possible values,
e.g., $\rho=\pm \sqrt{(1 - \phi^2)/R}$. The plots of $F$ and
$F_{SLC}$ for this case are shown against $R$ for different values
of $\phi$ in Fig. \ref{fig:F_onetap}. Figs. \ref{fig:F_unif} and
\ref{fig:F_onetap} point to similar behaviors of $F$ versus
$F_{SLC}$. Namely, $F$ becomes smaller than $F_{SLC}$ as $\phi$
decreases for a range of $R$ values. At these $R$ values, the
provable lower bound $I'(X; Y)$ is apparently tighter than the
SLC, with respect to the SIR.

\section{Bounding $F$}\label{sec:LB_MMI}
Exact computation of $F$ in general is not easy, especially when
$L$ goes to infinity. We thus resort to bounding $F$ with
expressions that can easily be computed. An upper bound on $F$
will provide a lower bound on $I'(X; Y)$ and thus on $I(X; Y)$.
Lower bounds on $F$ are also derived to see if they can get
smaller than $F_{SLC}$. If so, this would mean $I'(X; Y) = \log 2
- F$ is larger than $C_{SLC}(R) = \log 2 - F_{SLC}$, i.e., our
bound is tighter than the SLC.

\subsection{Simple Bounds}
Since $
 \log \Big( 1 + 2 \cosh(2R \rho^+) e^{-2 \phi \sqrt{R}
\tau} + e^{-4 \phi \sqrt{R} \tau - 4R} \Big)  
$ is convex in $\rho^+$, its integral function with respect to $\tau$, $\mathrm{E}_{\tau} \Big[ \frac{1}{2} \log
\Big( 1 + 2 \cosh(2R \rho^+) e^{-2 \phi \sqrt{R} \tau} + e^{-4
\phi \sqrt{R} \tau - 4R} \Big) \Big]$, is also convex in
$\rho^+$. Moreover, this function increases as $\rho^+$
increases. Accordingly, we can develop bounds on $F$. The first
simple upper bound is
\begin{eqnarray}
F^{u1} & \triangleq & T \left( |\rho|_{\max}, \theta \right) \Big\vert_{\theta = \sigma_{\rho}} \label{F^{u1}}
\end{eqnarray}
where, for a given $|\rho|_{\max}$, the function $T
\left(|\rho|_{\max}, \theta \right)$ represents a straight line
passing through two points of the function $\mathrm{E}_{\tau}
\Big[ \frac{1}{2} \log \Big( 1 + 2 \cosh(2R \theta) e^{-2 \phi
\sqrt{R} \tau} + e^{-4 \phi \sqrt{R} \tau - 4R} \Big) \Big]$
at $\theta=0$ and at $\theta=|\rho|_{\max}$. Note that
$|\rho|_{\max} \triangleq \max |\rho_i| = \sum_{k=1}^{L}
|d_{-k}|$ and $\sigma_{\rho}$ is the standard deviation of RV
$\rho$.

Similarly, $\mathrm{E}_{\tau} \left[ \frac{1}{2} \log \left( 1 + 2
\alpha e^{-2 \phi \sqrt{R} \tau} + e^{-4 \phi \sqrt{R} \tau - 4R}
\right) \right]$ is a concave and increasing function of $\alpha
\triangleq \cosh(2R \rho^+)$. Based on this property, we can
develop another upper bound.
\begin{eqnarray}
F^{u2} & \triangleq & \mathrm{E}_{\tau} \bigg[ \frac{1}{2}\log \Big\{ 1 + 2( s \sigma_{\rho} + 1) e^{ - 2\phi \sqrt{R} \tau  - 2R} \nonumber \\
& & \quad + e^{ - 4\phi \sqrt{R} \tau  - 4R}  \Big\} \bigg] \label{F^{u2}}
\end{eqnarray}
where $s = \left( \cosh(2R|\rho|_{\max}) - 1 \right) /
|\rho|_{\max}$, the slope of a straight line connecting two points
$(0, 1)$ and $(|\rho|_{\max}, \cosh(2R|\rho|_{\max}))$.

A lower bound on $F$ can also be obtained that can help shed
lights on how tight the upper bounds on $F$ are. Using the
convexity of $\mathrm{E}_{\tau} \Big[ \log \Big( 1 + e^{-2 R
\rho} e^{-2 \phi \sqrt{R} \tau - 2R} \Big) \Big]$ in $\rho$,
the simple lower bound of $F$ is
\begin{eqnarray}
F^l & \triangleq & \mathrm{E}_{\tau} \negthickspace \left[ {\frac{1}{2}\log \left\{ {1 + 2e^{ - 2\phi \sqrt R \,\tau  - 2R}  + e^{ - 4\phi \sqrt R \,\tau  - 4R} } \right\}} \right] \negthickspace . \label{F^{l}}
\end{eqnarray}
Detailed derivations of (\ref{F^{u1}}), (\ref{F^{u2}}), and
(\ref{F^{l}}) are given in Appendix \ref{simple_bounds}.

\subsection{Tightened Bounds Based on Cluster Identification}

The above bounds can be tightened up by identifying clusters in
the Gaussian mixture $f_V(t)$. In practical ISI channels, $f_V(t)$
often consists of clusters. This is due to the fact that the
coefficient set $d_{-k}$'s typically contains a few dominating
coefficients plus many small terms. Assuming there are $M$
dominating coefficients among $d_{-k}$'s, we can let $\rho_k =
\lambda_n + \mu_i$ where $n=1,2,\ldots, 2^M$, $i=1,2,\ldots,
2^{L-M}$, and $k = (n-1)2^{L-M}  + i$. Since $X_k$ is an i.u.d.
RV, $\lambda$ and $\mu$ are independent so that $\sigma_{\rho}^2 =
\sigma_{\lambda}^2 + \sigma_{\mu}^2$ where $\sigma_{\lambda}^2$
and $\sigma_{\mu}^2$ denote the variance of RVs $\lambda$ and
$\mu$, respectively. Notice that $\lambda_n$ can be viewed as the
position of a specific cluster while $\mu_i$ points to a specific
Gaussian pdf out of $2^{L-M}$ Gaussian pdf's symmetrically
positioned around $\lambda_n$.

Therefore, assuming there are $2^M$ clusters of Gaussian pdfs, the
upper bound $F^{u1}$ can be tightened as
\begin{eqnarray}
  F_M^{u1} & \triangleq & 2^{ -M} \sum_{n = 1}^{2^M } {T_n \left( |\mu|_{\max}, \theta \right)} \Big\vert_{\theta=\sigma_{\mu}} \label{eq:F_M^{u1}_BPSK}
\end{eqnarray}
where, for a given $|\mu|_{\max}$, the function $T_n(|\mu|_{\max},
\theta)$ is a straight line that passes through the two
points of the convex function $\mathrm{E}_{\tau} \Big[
\frac{1}{2} \log \Big\{ 1 + 2\cosh \left( 2R \theta  \right)
e^{ -2 R \lambda_n } e^{ - 2\phi \sqrt{R} \tau  - 2R}  + e^{ -4 R
\lambda_n } e^{ - 4\phi \sqrt{R} \tau  - 4R}  \Big\} \Big]$
at $\theta = 0$ and $\theta=|\mu|_{\max}$, $\sigma_{\mu}$ is the
standard deviation of RV $\mu$ defined as $\sigma_{\mu} =
\sqrt{\sigma_{\rho}^2 - \sigma_{\lambda}^2}$, and $|\mu|_{\max} =
|\rho|_{\max} - |\lambda|_{\max}$.

Another form of tightened upper bound based on $F^{u2}$ is
obtained as
\begin{eqnarray}
F_M^{u2}  & \triangleq & 2^{ -M} \sum_{n=1}^{2^M } \mathrm{E}_{\tau} \bigg[ \dfrac{1}{2} \log \Big\{ 1 + 2\left( {s_M \sigma_{\mu}  + 1} \right) e^{ - 2R\lambda_n } \nonumber \\
& & \quad \times e^{ - 2\phi \sqrt{R} \tau  - 2R}  + e^{ - 4R\lambda_n } e^{ - 4\phi \sqrt{R}\tau  - 4R}  \Big\} \bigg]  \label{eq:F_M^{u2}_BPSK}
\end{eqnarray}
where $s_M = \left(\cosh(2R|\mu|_{\max})-1 \right) / |\mu|_{\max}$.

The lower bound $F^l$ can also be tightened similarly based on the
cluster identification:
\begin{eqnarray}
  F_M^l & \triangleq & 2^{ -(M-1)} \sum_{k = 1}^{2^{M - 1} } \mathrm{E}_{\tau} \bigg[ \dfrac{1}{2} \log \Big\{ 1 + 2\cosh \left( {2R\lambda_k^+ } \right) \nonumber \\
 & & \quad  \times e^{ - 2\phi \sqrt{R} \tau  - 2R}  + e^{ - 4\phi \sqrt{R} \tau  - 4R}  \Big\} \bigg]  \label{eq:F_M^{l}_BPSK}
\end{eqnarray}
where $\lambda_k^+$'s form the positive-half subset of
$\lambda_n$'s. Detail derivations of (\ref{eq:F_M^{u1}_BPSK}),
(\ref{eq:F_M^{u2}_BPSK}), and (\ref{eq:F_M^{l}_BPSK}) can be found
in Appendix \ref{tight_bounds}.

\subsection{Bounds for Complex Channels with the Quaternary Alphabet Inputs}\label{subsec:LB_Complex}
In the previous subsections, ISI coefficients and noise samples
are assumed to be real-valued with the channel inputs being the
binary phase shift keying (BPSK) signal. In this subsection, we
provide a complex-valued example along with the channel inputs
taken from a quadrature phase shift keying (QPSK) quaternary
alphabet, i.e., $X_k \in \Big\{ \sqrt{ \frac{P_X} {2} } (\pm 1 \pm j ) \Big\}$.
The extension to larger alphabets should be straightforward.

Denoting the real and imaginary parts of complex number $a$ by $a^{(r)}$ and $a^{(i)}$ respectively, i.e., $a =
a^{(r)} + j a^{(i)}$, and $m_i = \sum_{k=1}^{L} d_{-k}X_{k}$ for
$i=1,2,\ldots, 4^L$, the pdf's of complex random variables $V$
and $G$ are given as
\begin{eqnarray}
f_V(t) & = & 4^{-L} \sum_{i =1}^{4^L} \dfrac{1}{\pi \sigma_{N}^2 } \exp \left( - \dfrac{ \left| t - m_i \right |^2 } {\sigma_{N}^2 } \right)  \nonumber \\
& = & 4^{-L} \sum_{i =1}^{4^L}  \Bigg \{ \dfrac{1}{ \sqrt{ \pi \sigma_{N}^2 } } \exp \Bigg( - \dfrac{ \big( t^{(r)} - m_i^{(r)} \big )^2 } {\sigma_{N}^2 } \Bigg) \nonumber \\
& & \times \dfrac{1}{ \sqrt{ \pi \sigma_{N}^2 } } \exp \Bigg( - \dfrac{ \big( t^{(i)} - m_i^{(i)} \big )^2 } {\sigma_{N}^2 } \Bigg) \Bigg \} \nonumber \\
f_G(t) & = &  \dfrac{1}{ \pi  \sigma_{V}^2 } \exp \left( - \dfrac{ |t|^2 } { \sigma_{V}^2 } \right) \nonumber \\
& = &  \dfrac{1}{ \sqrt{ \pi  \sigma_{V}^2 } } \exp \left( - \dfrac{ \left( t^{(r)} \right)^2 } { \sigma_{V}^2 } \right) \dfrac{1}{ \sqrt{ \pi  \sigma_{V}^2 } } \exp \left( - \dfrac{ \left( t^{(i)} \right)^2 } { \sigma_{V}^2 } \right) . \nonumber
\end{eqnarray}
Then, for the SLC, we write \sublabon{equation}
\begin{eqnarray}
F_{SLC} & \triangleq & \log 4 - C_{SLC} (R) \nonumber \\
&  = & 2  \int_{ - \infty }^{\infty}  \dfrac{e^{ - \tau^2 } } { \sqrt{\pi} }  \log \left \{ {1 + e^{ - 2\sqrt{2R} \tau  - 2R} } \right\} d\tau \nonumber \\
& = & 2  \mathrm{E}_{\tau} \left[ \log \left \{ {1 + e^{ - 2\sqrt{2R} \tau  - 2R} } \right\} \right] \\
&  = & 2  \mathrm{E}_{\tau} \left[ \dfrac{1}{2} \log \left \{ {1 + 2e^{ - 2\sqrt{2R} \tau  - 2R} + e^{ - 4\sqrt{2R} \tau  - 4R} } \right\} \right] \nonumber \\
\end{eqnarray}\sublaboff{equation}
where $\mathrm{E}_{\tau} \left[ p(\tau) \right] = \int_{-\infty}^{\infty}  \pi^{-1/2} e^{-\tau^2}  p(\tau) d\tau$.

The function $F$ is given as \sublabon{equation}
\begin{eqnarray}
F & \triangleq & 4^{-L} \sum_{i=1}^{4^L} \bigg( \mathrm{E}_{\tau} \left[ \log \left \{ 1 + e^{-2\sqrt{2} R \rho^{(r)}_i} e^{-2 \phi \sqrt{2R} \tau - 2R}  \right \} \right] \nonumber \\
& & +\mathrm{E}_{\tau} \left[ \log \left \{ 1 + e^{-2\sqrt{2} R \rho^{(i)}_i} e^{-2 \phi \sqrt{2R} \tau - 2R}  \right \} \right] \bigg) \nonumber \\
& = & 2 \mathrm{E}_{\rho^{(r)}, \tau} \left[ \log \left \{ 1 + e^{-2\sqrt{2} R \rho^{(r)}} e^{-2 \phi \sqrt{2R} \tau - 2R}  \right \} \right]  \label{eq:F_1_QPSK} \\
& = & 4^{ - (L - 1)} \sum_{k = 1}^{4^{L - 1} } 2 \mathrm{E}_{\tau} \bigg[  \dfrac{1}{2} \log \Big\{ 1 + 2\cosh \big( 2\sqrt{2} R\rho_k^{(r)+}  \big) \nonumber \\
& & \quad \times e^{ - 2\phi \sqrt{2R} \tau  - 2R}  + e^{ - 4\phi \sqrt{2R} \tau  - 4R}  \Big\}  \bigg] \nonumber \\ 
& = & 2 \mathrm{E}_{\rho^{(r)+}, \tau} \bigg[ \frac{1}{2}\log \Big\{ 1 + 2 \cosh \big( 2\sqrt{2} R \rho^{(r)+}   \big) e^{ - 2\phi \sqrt{2R} \tau  - 2R} \nonumber \\
& & \quad + e^{ - 4\phi \sqrt{2R} \tau  - 4R}  \Big \} \bigg] \label{eq:F_2_QPSK}
\end{eqnarray}\sublaboff{equation}where $\rho_i^{(r)} \triangleq m_i^{(r)} /\sqrt{P_X}$,
$\rho_i^{(i)} \triangleq m_i^{(i)} /\sqrt{P_X}$, and
$\rho_k^{(r)+}$'s and $\rho_k^{(i)+}$'s denote the positive-half
subset of $\rho_i^{(r)}$'s and $\rho_i^{(i)}$'s respectively. The
equality (\ref{eq:F_1_QPSK}) holds because the pdf of $\rho^{(r)}$ is identical to the pdf of $\rho^{(i)}$.

Then, the upper bound based on $F^{u1}$ can be derived in a similar
way as
\begin{eqnarray}
  F_M^{u1} & \triangleq &   4^{-M} \sum_{n = 1}^{4^M } 2 T^{(r)}_n \left( |\mu^{(r)}|_{\max}, \theta   \right)\Big\vert_{ \theta = \frac{\sigma_{\mu}}{ \sqrt{2}} } \label{eq:F_M^{u1}_QPSK}
\end{eqnarray}
where, for a given $|\mu^{(r)}|_{\max}$,
$T^{(r)}_n(|\mu^{(r)}|_{\max}, \theta)$ denotes a straight line
that passes through the two points of the function $\mathrm{E}_{\tau} \Big[
\frac{1}{2} \log \Big\{ 1 + 2\cosh \left( {2\sqrt{2} R \theta }
\right) e^{ -2\sqrt{2} R \lambda^{(r)}_n } e^{ - 2\phi \sqrt{2R}
\tau  - 2R}  + e^{ -4\sqrt{2} R \lambda^{(r)}_n } e^{ - 4 \phi
\sqrt{2R} \tau  - 4R}  \Big\} \Big]$ at $\theta = 0$ and at
$\theta=|\mu^{(r)}|_{\max}$. Note that $|\mu^{(r)}|_{\max}=|\mu|_{\max}/\sqrt{2}$ and the variance of $\mu^{(r)}$ is equal to $\sigma_{\mu}^2/2$ since the pdfs of $\lambda^{(r)}$ and
$\mu^{(r)}$ are identical to the pdfs of $\lambda^{(i)}$ and
$\mu^{(i)}$, respectively.

A second upper bound on $F$ is given as
\begin{eqnarray}
F_M^{u2}  & \triangleq &  4^{ -M} \sum_{n=1}^{4^M } 2 \mathrm{E}_{\tau} \bigg[ \dfrac{1}{2} \log \bigg\{ 1 + 2\bigg(  \dfrac{s^{(r)}_M \sigma_{\mu}}{\sqrt{2}}   + 1 \bigg) e^{ - 2\sqrt{2} R\lambda^{(r)}_n } \nonumber \\
& & \quad \times e^{ - 2\phi \sqrt{2R} \tau  - 2R}  + e^{ - 4\sqrt{2} R\lambda^{(r)}_n } e^{ - 4\phi \sqrt{2R}\tau  - 4R}  \bigg\} \bigg] 
\label{eq:F_M^{u2}_QPSK}
\end{eqnarray}
where $s^{(r)}_M = \left(\cosh(2 \sqrt{2} R|\mu^{(r)}|_{\max})-1
\right) / |\mu^{(r)}|_{\max}$.

\begin{figure*}[!t]
\centering 
\includegraphics[width=14.0cm]{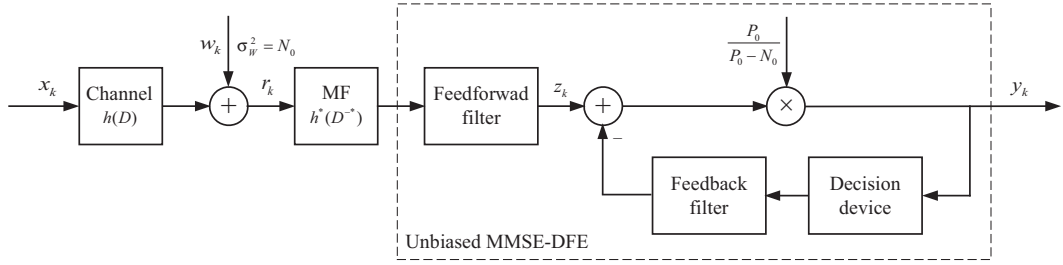}
\caption{System Model of ISI channels.}
\label{fig:ISI_Model}
\end{figure*}

Finally, a lower bound to $F$ can be shown to be
\begin{eqnarray}
F_M^{l}  & \triangleq &  4^{ -M} \sum_{k=1}^{4^M /2}  \mathrm{E}_{\tau} \bigg[ \dfrac{1}{2} \log \Big\{ 1 + 2 \cosh \big( 2\sqrt{2} R\lambda^{(r)+}_k \big) \nonumber \\
& & \quad \times e^{ - 2\phi \sqrt{2R} \tau  - 2R}  + e^{ - 4\phi \sqrt{2R}\tau  - 4R}  \Big\} \bigg] \label{eq:F_M^{l}_QPSK}
\end{eqnarray}
where $\lambda^{(r)+}_k$'s form the
positive-half subset of $\lambda^{(r)}_n$'s.

\section{Application to ISI Channels and Numerical Examples}\label{sec:Numerical Results}

\subsection{The ISI Channel and MMSE-DFE}
Fig. \ref{fig:ISI_Model} shows the discrete-time equivalent system
model of the finite-ISI channel with the infinite-length
feedforward filter of the unbiased MMSE-DFE preceded by the matched
filter (MF) for the channel. The discrete-time MF output of Fig.
\ref{fig:ISI_Model} is identical to the baud-rate sampled output
of the continuous-time MF applied to the continuous-time channel,
under the assumption that the channel is strictly limited to the
Nyquist band.

We also assume that the receiver knows the $D$-transform
of the finite-ISI channel response, $h(D)$, $x_k$ is an i.u.d.
input sequence and $w_k$ is additive white Gaussian noise (AWGN)
with variance $\sigma_W^2 = N_0$. Furthermore, $r_k$ is the
channel output sequence, $z_k$ is the output sequence of the
infinite-length MMSE-DFE feedforward filter and $y_k$ is the unbiased MMSE-DFE output after ideal postcursor ISI cancellation.

Denoting $X = x_0$, $X_k = x_k$, and $Y = y_0$, the output of the
the unbiased MMSE-DFE with ideal feedback \cite{Cioffi95} is 
\begin{eqnarray}
Y = X + \sum_{k=1}^{\infty} d_{-k} X_k + N = X + S + N = X + V \nonumber
\end{eqnarray}
where $N$ is the Gaussian noise sample observed at the DFE forward
filter output and $d_{-k} X_k$ is the precursor ISI sequence. Note
we are assuming stationary random processes. It is well-known that
the $D$-transform of the precursor ISI taps $d_{-k}$ is given by
\cite{Cioffi95}
\begin{eqnarray}
d(D) = \dfrac{N_0}{P_0 - N_0} \left(1 - \dfrac{1}{g^*(D^{-*})} \right)
\end{eqnarray}
where $P_0$ is such that $\log P_0 = \frac{1}{2\pi}
\int_{-\pi}^{\pi} \log R_{ss}(e^{-j\theta}) d\theta$ and
$g^*(D^{-*})$ is obtained from spectral factorization: $R_{ss}(D) = P_X
R_{hh}(D) + N_0 = P_0 g(D) g^*(D^{-*})$ with $R_{hh}(D) =
h(D)h^*(D^{-*})$. Notice that a convenient numerical spectral
factorization algorithm exists for recursively computing the
coefficients of $g^*(D^{-*})$ \cite{Messerschmitt73},
\cite{Forney98}.

Accordingly, the variances of $V$, $N$, and $S$
are given as
\begin{eqnarray}
 \sigma _V^2 & = & \dfrac{{P_X N_0 }}{{P_0  - N_0 }} \nonumber \\
 \sigma _N^2  & = & \dfrac{{P_X P_0 N_0 }}{{2\pi \left( {P_0  - N_0 } \right)^2 }}\int_{ - \pi }^\pi  {\frac{{R_{hh} (e^{ - j\theta } )}}{{R_{hh} (e^{ - j\theta } ) + N_0 / P_X }}d\theta }  \nonumber \\
 \sigma _S^2  & = & \sigma _V^2  - \sigma _N^2 . \nonumber
\end{eqnarray}
We can obtain $|\rho|_{\max}$ by the absolute summation of the
inverse $D$-transform of $d(D)$ if the feedforward filter of
MMSE-DFE is stable, i.e., $\sum_{k=1}^{\infty} | d_{-k} | <
\infty$. Let us first consider the case where $d(D)$ has $P$
multiple first-order poles, $p_i$ for $i= 1,2,\ldots, P$. Then,
$|\rho|_{\max}$ can be obtained by the partial fraction method
since $d(D)$ is a rational function. In other words,
the inverse $D$-transform of individual fraction terms can be
found and then added together to form $d_{-k}$. Denoting $a(D) =
\frac{1}{{g^* (D^{ - *} )}} = \sum_{i = 1}^P {\frac{{c_i }}{{1 -
p_i D^{ - 1} }}}$, the sequence $a_{-k}$ is given as $a_{ - k}  =
\sum_{i = 1}^P {c_i p_i^k }$. Therefore,
\begin{eqnarray}
  \left| \rho  \right|_{\max } & = & \dfrac{1}{\sqrt{P_X}} \sum_{k=1}^{\infty} \left| d_{-k} X_k
\right|  = \sum_{k=1}^{\infty} \left| d_{-k} \right| \nonumber \\
& = & \dfrac{{N_0 }}{{ \left( {P_0  - N_0 } \right)}}\left( {\sum\limits_{k = 1}^\infty  {\left| {a_{ - k} } \right|} } \right) \nonumber \\
& = & \dfrac{{N_0 }}{{ \left( {P_0  - N_0 } \right)}}\left( {\sum\limits_{k = 1}^\infty  {\left| {\sum\limits_{i = 1}^P {c_i p_i^k } } \right|} } \right) \nonumber \\
  & \leq & \dfrac{{N_0 }}{{ \left( {P_0  - N_0 } \right)}}\left( {\sum\limits_{i = 1}^P {\sum\limits_{k = 1}^\infty  {\left| {c_i p_i^k } \right|} } } \right) \nonumber \\
  &  = & \dfrac{{N_0 }}{{ \left( {P_0  - N_0 } \right)}}\left( {\sum\limits_{i = 1}^P {\dfrac{{\left| {c_i p_i } \right|}}{{1 - \left| {p_i } \right|}}} } \right). \label{eq:rho_max1}
\end{eqnarray}
The upper bound of $|\rho|_{\max}$ can be also tightened by
identifying the first $K$ dominant taps:
\begin{eqnarray}
  \left| \rho  \right|_{\max }  & = & \dfrac{{N_0 }}{{\left( {P_0  - N_0 } \right)}}\left( {\sum\limits_{k = 1}^\infty  {\left| {\sum\limits_{i = 1}^P {c_i p_i^k } } \right|} } \right) \nonumber \\
   & = & \dfrac{{N_0 }}{{ \left( {P_0  - N_0 } \right)}} \left( {\sum\limits_{k = 1}^K {\left| {\sum\limits_{i = 1}^P {c_i p_i^k } } \right|}  + \sum\limits_{k = K + 1}^\infty  {\left| {\sum\limits_{i = 1}^P {c_i p_i^k } } \right|} } \right) \nonumber \\
   & \leq & \dfrac{{N_0 }}{{ \left( {P_0  - N_0 } \right)}} \left( {\sum\limits_{k = 1}^K {\left| {\sum\limits_{i = 1}^P {c_i p_i^k } } \right|}  + \sum\limits_{i = 1}^P {\sum\limits_{k = K + 1}^\infty  {\left| {c_i p_i^k } \right|} } } \right) \nonumber \\
  & = & \sum_{k=1}^{K} | d_{-k} | + \dfrac{{N_0 }} {{ \left( {P_0  - N_0 } \right)}} \left( \sum\limits_{i = 1}^P { \dfrac{ {\left| {c_i p_i^{K + 1} } \right|} } { {1 - \left| {p_i } \right|} } }  \right). \label{eq:rho_max2}
\end{eqnarray}
For the case of the multiple-order poles of $d(D)$, the upper
bound of $|\rho|_{\max}$ can be also obtained in a similar way
using the triangle inequality $|a+b| \leq |a| + |b|$.

The SIR or the i.u.d. capacity (bits/channel use) for any finite-ISI
channel corrupted by Gaussian noise is given \cite{Gallager68} as
\begin{eqnarray}
\mathrm{SIR} & \triangleq & \lim_{N \rightarrow \infty} \dfrac{1} {2 N + 1} I \left( \{x_k \}_{-N}^{N}; \{r_k \}_{-N}^{N} \right) \nonumber \\
& \geq & \lim_{N \rightarrow \infty} \dfrac{1} {2 N + 1} I \left( \{x_k \}_{-N}^{N}; \{z_k \}_{-N}^{N} \right) \label{eq:C_L1} \\
& \geq &  I \left( x_0; z_0 | \{x_k \}_{-\infty}^{-1} \right) \label{eq:C_L1A} \\
& = &  I \left( X ; Y \right) \label{eq:C_L2}
\end{eqnarray}
where $\{u_k\}_{N_1}^{N_2} = \{u_k, k = N_1, N_1+1, \ldots, N_2
\}$. The inequality in (\ref{eq:C_L1}) holds due to the data
processing theorem (equality holds if the MMSE-DFE feedforward
filter is invertible). The inequality of
(\ref{eq:C_L1A}) can be obtained by applying the chain rule of
mutual information and assuming stationarity \cite{Shamai96}.
The equality (\ref{eq:C_L2}) is valid because known post-cursor ISI can simply be subtracted out
without affecting capacity.

\begin{figure}[!t]
\centering 
\includegraphics[width=9cm]{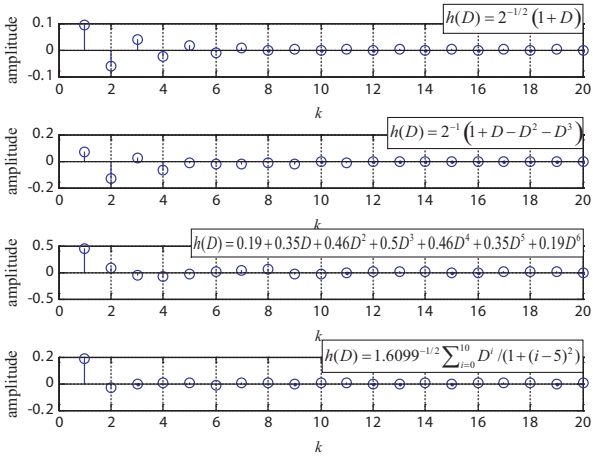}
\caption{First 20 Precursor taps after unbiased MMSE-DFE at SNR=10 dB for four example channels.}
\label{fig:Precursor}
\end{figure}

\begin{figure}[!t]
\centering 
\subfigure[] 
{
    \includegraphics[width=8.5cm]{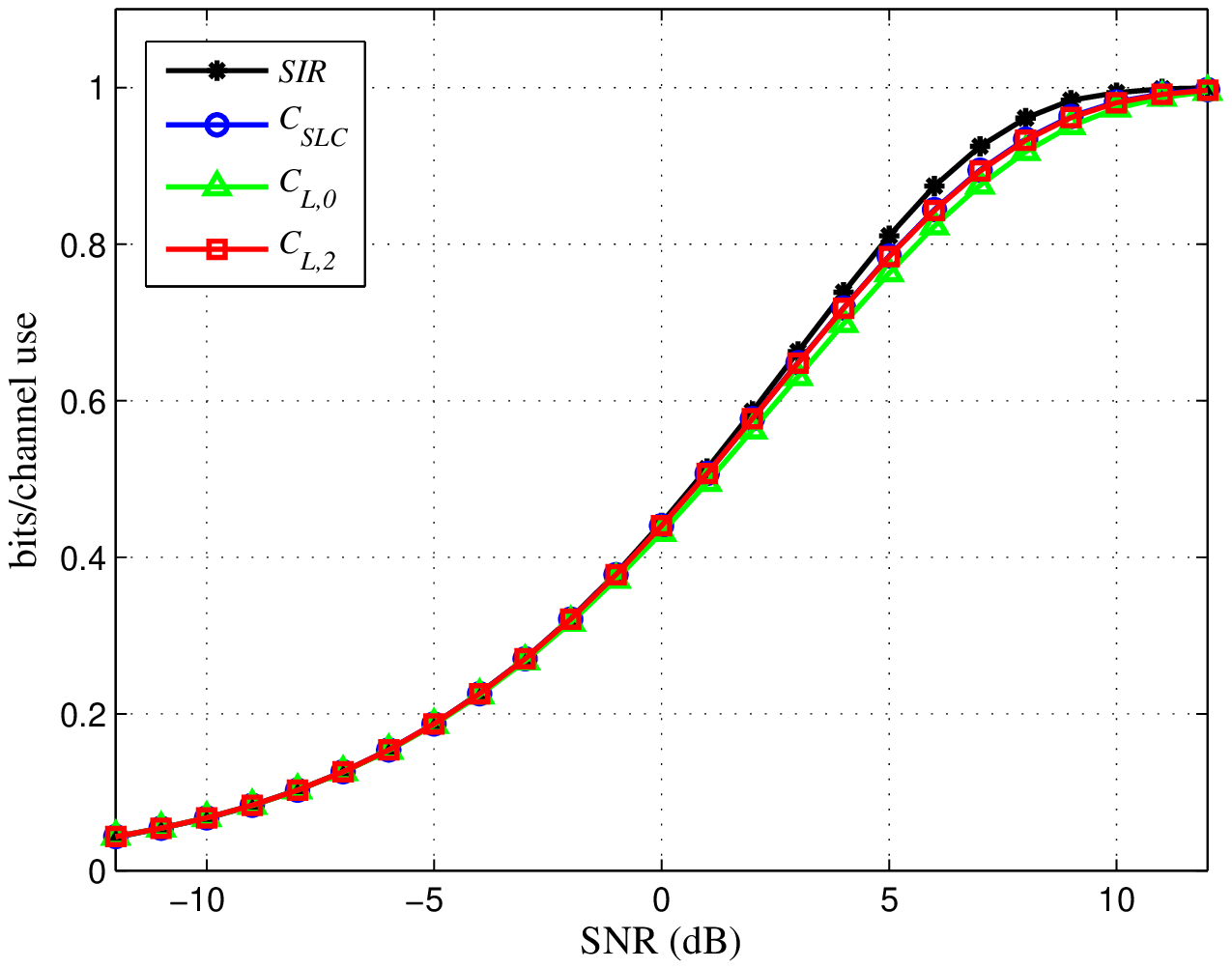}
}
\subfigure[] 
{
    \includegraphics[width=8.5cm]{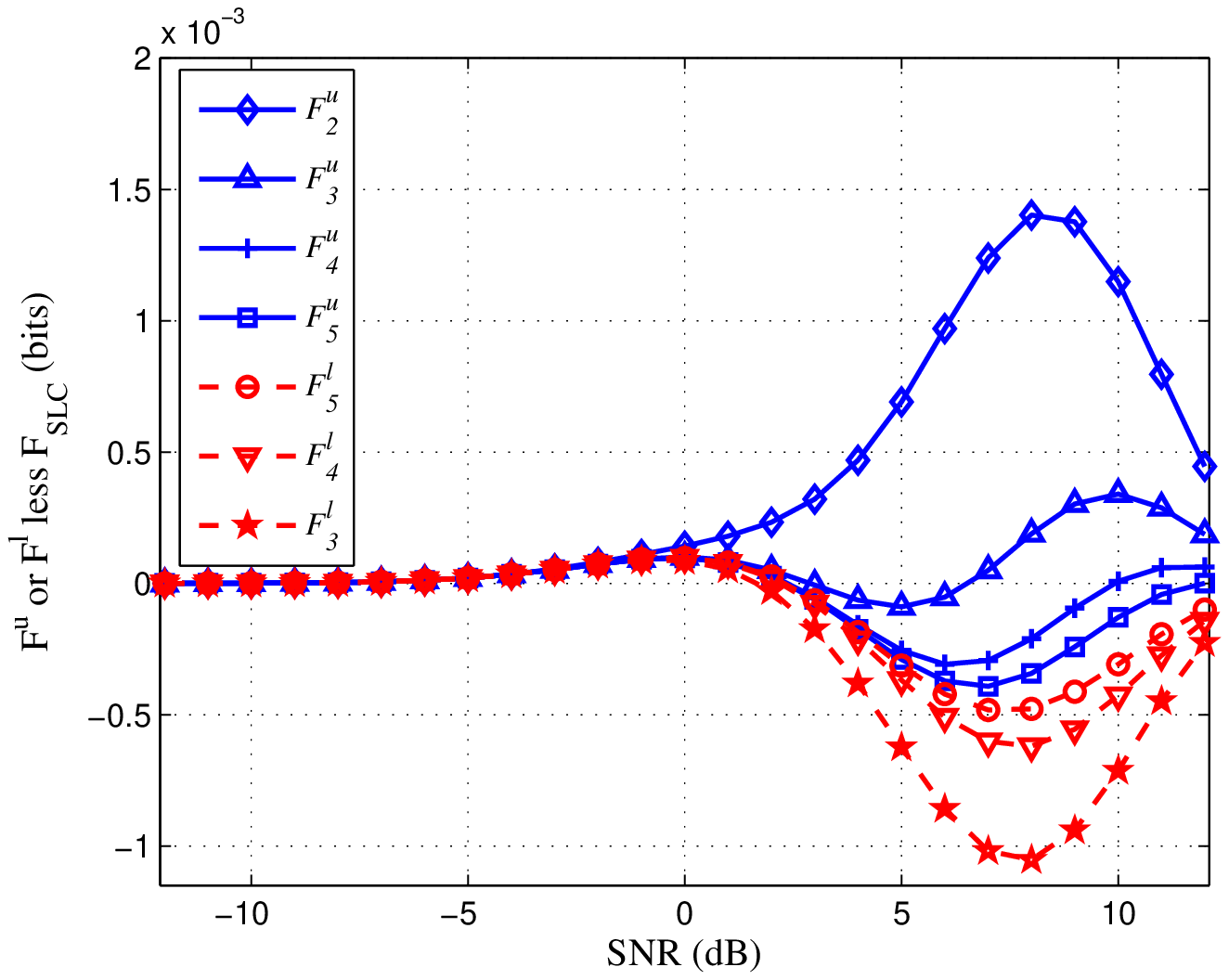}
} \caption{Example channel $h(D) = 2^{-1/2}( 1 + D )$ with BPSK inputs (a) SIR, SLC and the new lower bounds as functions of SNR (b)
Upper and lower bounds of $F$, for different M, less $F_{SLC}$, plotted against SNR.}
\label{fig:example1}
\end{figure}

\begin{figure}[!t]
\centering 
\subfigure[] 
{
    \includegraphics[width=8.5cm]{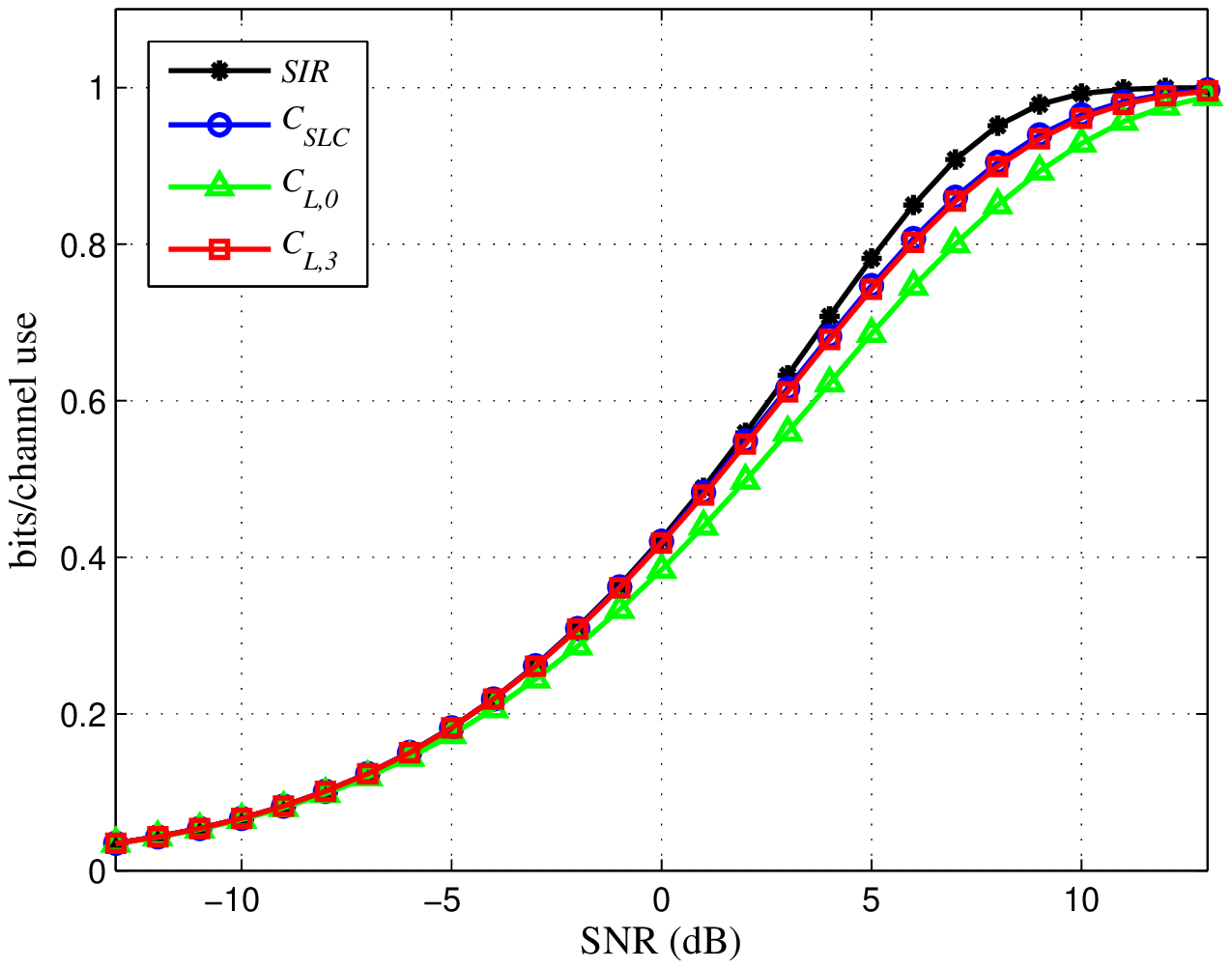}
}
\subfigure[] 
{
    \includegraphics[width=8.5cm]{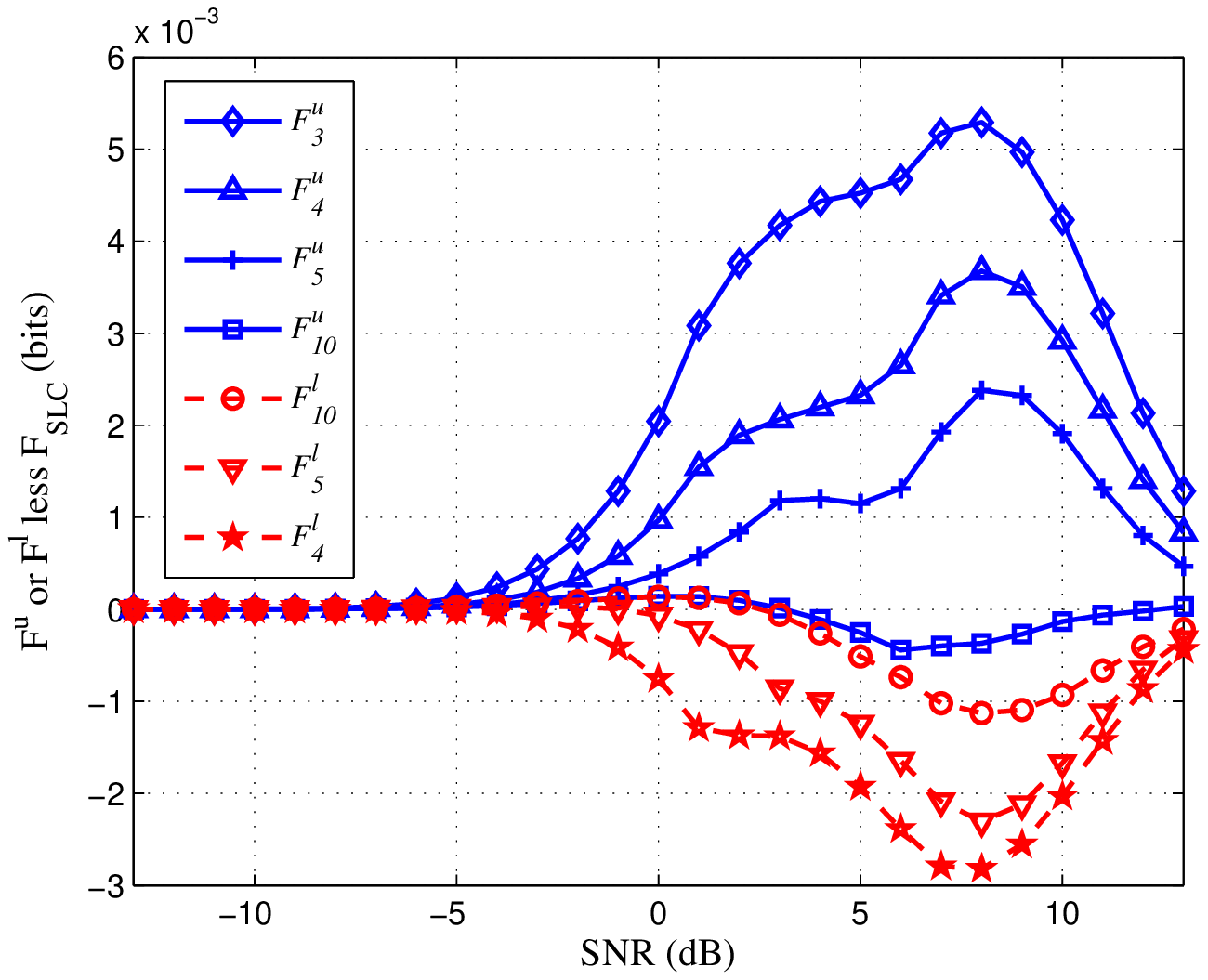}
} \caption{Example channel $h(D) =  2^{-1} (1 + D - D^2 - D^3)$ with BPSK inputs (a) SIR, SLC and the new lower bounds as functions of SNR (b)
Upper and lower bounds of $F$, for different M, less $F_{SLC}$, plotted against SNR.}
\label{fig:example2}
\end{figure}

\subsection{Numerical Results}
Now, let us examine the particular ISI channels,  $h(D) = 2^{-1/2}
(1 + D)$, $h(D) = 2^{-1} (1 + D - D^2 - D^3)$ and $h(D)  = 0.19 +
0.35D + 0.46D^2 + 0.5D^3 + 0.46D^4 + 0.35D^5 + 0.19D^6$, which are
well-known and previously investigated in \cite{Arnold01},
\cite{Shamai91}, \cite{Shamai96}, and $h(D) = 1.6099^{-1/2}
\sum_{i=0}^{10} D^i / (1+(i-5)^2)$, which was considered in
\cite{Sadeghi09}. The first 20 precursor ISI tap values are
computed and shown in Fig. \ref{fig:Precursor} for these example
channels. In addition, we consider a complex-valued partial
response channel: $h(D) = 2^{-1} \left\{ (1+j) + (1-j)D\right\} $.
The channel inputs are binary, except the complex-valued channel
for which the inputs are assumed quaternary.

\begin{figure}[!t]
\centering 
\subfigure[] 
{
    \includegraphics[width=8.5cm]{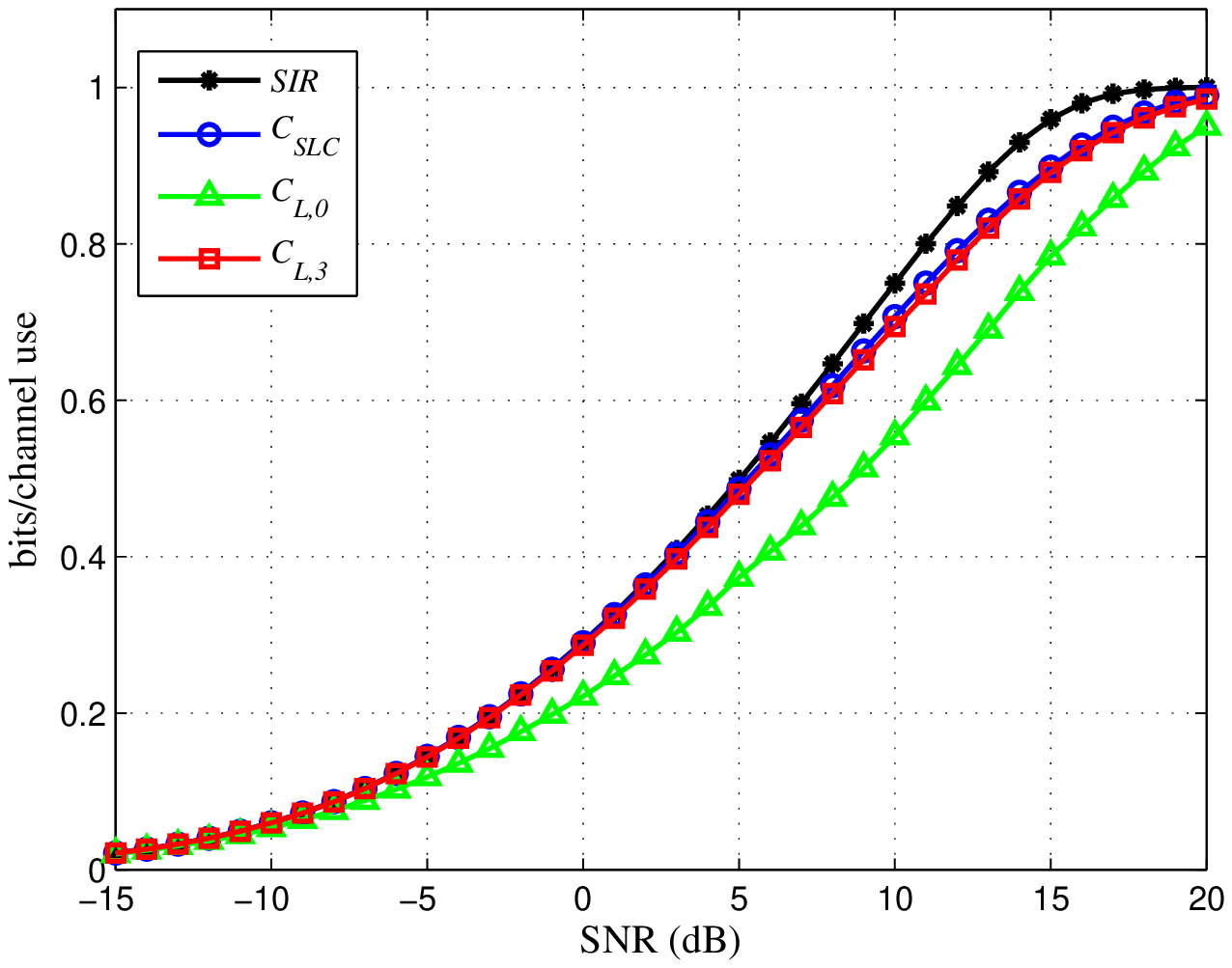}
}
\subfigure[] 
{
    \includegraphics[width=8.5cm]{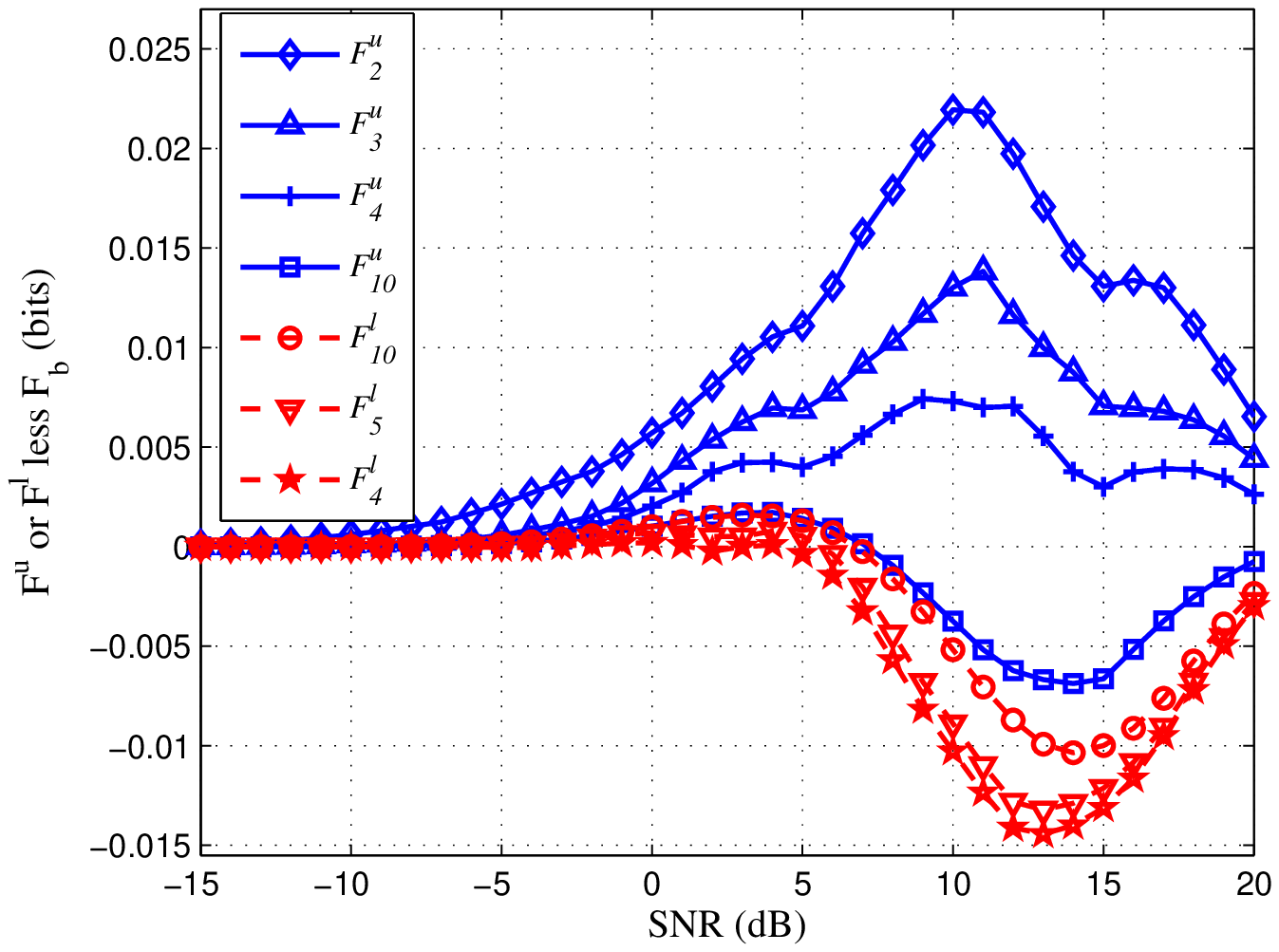}
} \caption{Example channel $h(D) = 0.19 + 0.35D +
0.46D^2 + 0.5D^3 + 0.46D^4 + 0.35D^5 + 0.19D^6$ with BPSK inputs
(a) SIR, SLC and the new lower bounds as functions of SNR (b)
Upper and lower bounds of $F$, for different M, less $F_{SLC}$, plotted against SNR.}
 \label{fig:example3}
\end{figure}
\begin{figure}[!t]
\centering 
\subfigure[] 
{
    \includegraphics[width=8.5cm]{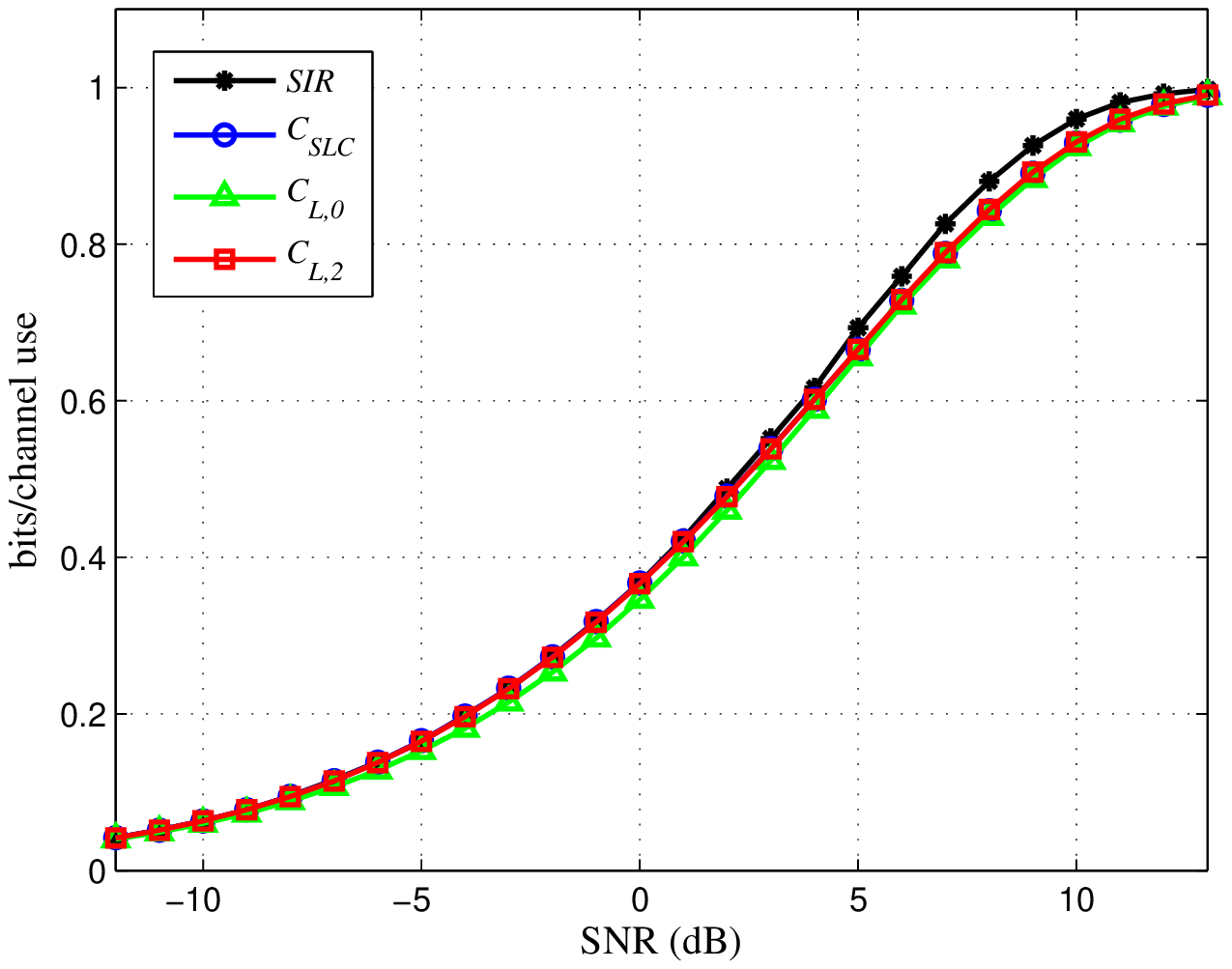}
}
\subfigure[] 
{
    \includegraphics[width=8.5cm]{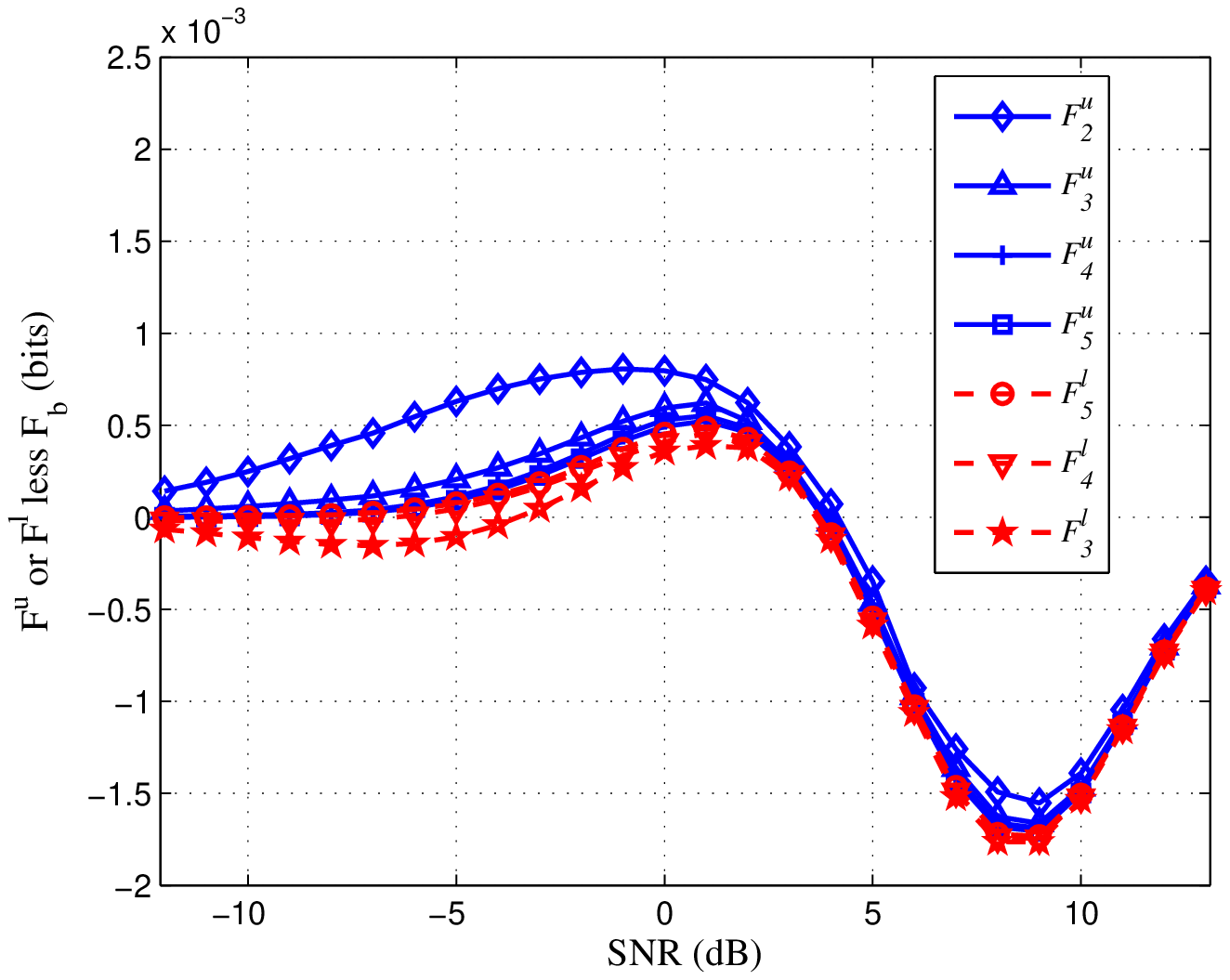}
} \caption{Example channel $h(D) = 1.6099^{-1/2} \sum_{i=0}^{10} D^i / (1+(i-5)^2)$ with BPSK inputs (a) SIR, SLC and the new lower bounds as functions of SNR (b)
Upper and lower bounds of $F$, for different M, less $F_{SLC}$, plotted against SNR.}
 \label{fig:example4}
\end{figure}

\begin{figure}[!t]
\centering 
\subfigure[] 
{
    \includegraphics[width=8.5cm]{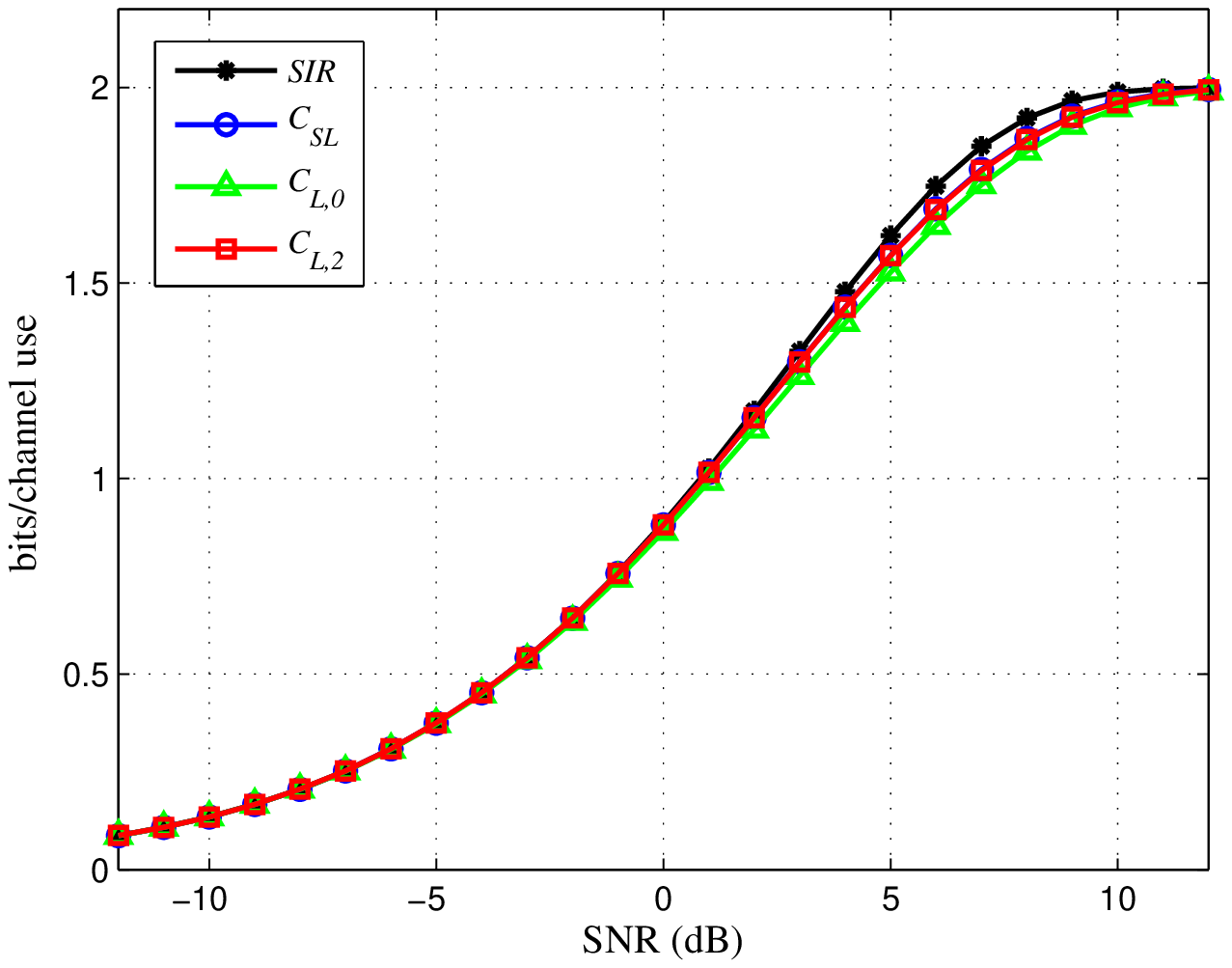}
}
\subfigure[] 
{
    \includegraphics[width=8.5cm]{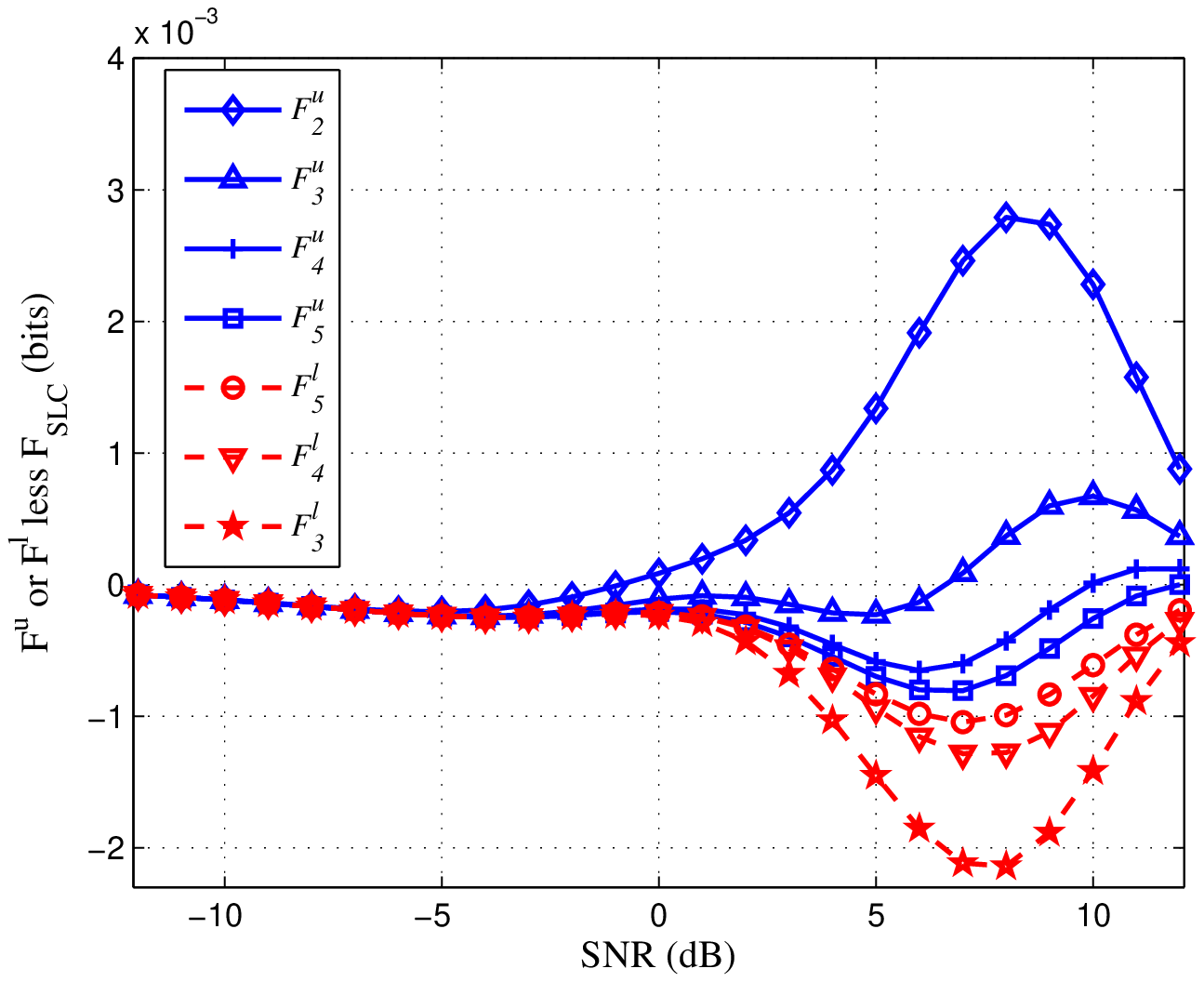}
} \caption{Example channel $h(D) =  2^{-1} \left\{ (1+j) + (1-j)D\right\}$ with QPSK inputs (a) SIR, SLC and the new lower bounds as functions of SNR (b)
Upper and lower bounds of $F$, for different M, less $F_{SLC}$, plotted against SNR.}
 \label{fig:example5}
\end{figure}

Since the infinite-length MMSE-DFE is used, i.e., $L=\infty$, the probability distribution of $\rho$ is not available generally.
Hence the lower bounds $C_{L1,M} = \log 2 - F_M^{u1}$ and
$C_{L2,M} = \log 2 - F_M^{u2}$ along with $C_{SLC} = \log 2 - F_{SLC}$
are considered as functions of $\mathrm{SNR} = P_X / N_0$ for
different values of $M$. When no clustering is used, we set $M =
0$. In computing $|\rho|_{\max}$ (and thus $|\mu|_{\max}$), which was needed
to calculate $F_M^{u1}$ or $F_M^{u2}$, we were able to run
numerical recursive spectral factorization to find all
non-negligible $d_{-k}$ coefficients relatively quickly for all
channels considered, without resorting to the bounds of
(\ref{eq:rho_max1}) or (\ref{eq:rho_max2}). We observed that the
lower bounds, $C_{L1,M}$ and $C_{L2,M}$, produced similar results,
so only $C_{L1,M}$ were chosen and plotted as $C_{L,M}$ in
Figs. \ref{fig:example1} through \ref{fig:example5}. The SIR of each
channel is also obtained using the simulation-based approach
\cite{Arnold01}, \cite{Sharma01}, \cite{Pfister01}.

For each capacity figure, we first plotted the SIR and $C_{SLC}$.
We then plotted $C_L$ for $M=0$ and then another $C_L$ by choosing
an $M$ value for which the $C_L$ bound is almost as tight as the
$C_{SLC}$ conjecture (this is why the $C_{SLC}$ curve is almost
overwritten and indistinguishable in some figures). We also show
for each channel how the upper and lower bounds of $F$ close on
each other as $M$ increases. The bounds on $F$ are shown with
$F_{SLC}$ subtracted from them in the part (b) figure. In this way, it should be clear
that for those SNR values where $F^u - F_{SLC}$ becomes less than
zero eventually, $F$ is less than $F_{SLC}$, guaranteeing that
$I'(X:Y)=\log 2-F$ is larger than $C_{SLC}$. In fact, it can be
seen from the part (b) plots of Figs. \ref{fig:example1} through \ref{fig:example4} that this is true for the high SNR range corresponding to all rates higher than roughly 0.6 in each of the first four channels
considered, although the difference $F - F_{SLC}$ is small (note the very small scale of the vertical axis in each part-b figure).
In the fifth example shown in Fig. \ref{fig:example5},
$F^u - F_{SLC}$ is actually less than zero at all SNR values, meaning that for this channel 
our bound is tighter than the SLC in the entire SNR range. 

A significant implication here is that whenever 
$I'(X:Y)=\log 2-F$ is larger than $C_{SLC}$, there is an  
assurance that the SLC holds true. The
curves of $F^l-F_{SLC}$ for different $M$ values also provide a
detailed picture of how large $M$ should be in order for $C_L$ to
get close enough to the $C_{SLC}$.

Note that the computational load for evaluating the integral of
(\ref{eq:F_M^{u1}_BPSK}) and (\ref{eq:F_M^{u2}_BPSK}) to obtain
the bound depends exponentially on $M$, the number of clusters in
the pdf $f_V(t)$. The computational load in computing the dominant
precursor ISI taps and their magnitude sum is minimal. The results
summarized in the figures indicate that in each channel
considered, a relatively small value of $M$ (and thus a reasonably
low computational load) yields a bound as tight as the SLC. As a
case in point, comparison of Fig. \ref{fig:example4} with the
results of \cite{Sadeghi09} (Figure 6 of \cite{Sadeghi09},
specifically) gives a good idea on the usefulness of an easily
computable bound such as the one presented here. At a rate 0.9,
for example, one can observe from a close examination of Fig. 6 of
\cite{Sadeghi09} that the lower bound of \cite{Sadeghi09}
approaches the SIR within about 0.88 dB with 2 iterations, which
would require basically running the BCJR algorithm twice on a
reduced channel trellis of 64 states. In contrast, our bound based
on just two clusters is about 0.84 dB away from the SIR at the
same rate, as estimated from Fig. \ref{fig:example4}. This bound
requires computation of $2^2=4$ single-dimensional integrals, the complexity of which amounts to virtually nothing relative
to that associated with two BCJR simulation runs in the method of
\cite{Sadeghi09}. The simulation-based bound of \cite{Sadeghi09}
does narrow the gap to about 0.65 dB with five iterations, but at
the expense of much more computational time.

We stress that the value of the simulation-based SIR estimation methods is not in their ability to provide easily
obtained bounds; rather they play a critical role in estimating the SIR (or capacity) with a very high accuracy, given the ample computational resources. As for providing convenient and easily computed SIR estimates or bounds, the need for analytically evaluated bounds such as the one developed in this paper continues to be high. In particular, the question remains as to how does the proposed method perform on very long ISI channels with no dominant taps. A good example of this type of channel is the indoor wireless channel in a highly scattered multi-path setting. Unfortunately, our analysis indicates the lower bounds developed here are not very effective in this case, with their gaps to the SIR bigger than that of the SLC when computational loads are kept at reasonable levels. Easily-computed tight bounds for this type of channel remain elusive.

\section{Conclusion}\label{sec:Conclusion}
In this paper, we derived a lower bound to the SIR of the ISI
channel driven by discrete and finite-amplitude inputs. The
approach taken was to introduce a ``mismatched" mutual information
function that acts as a lower bound to the symmetric information
rate between the channel input and the ideal-feedback MMSE-DFE
filter output. This function turns out to be tighter than the
Shamai-Laroia conjecture for a practically significant range of
SNR values for some example channels. We then further
lower-bounded this function by another function that can be
evaluated via numerical integration with a small computational
load. The final computation also requires finding a few large
precursor ISI tap values as well as the absolute sum of the
remaining ISI terms, which can be done easily. The final lower
bounds are demonstrated for a number of well-known finite-ISI
channels, and the results indicate that the new bounds computed at
a fairly low computational load are as tight as the SLC.

\appendices
\section{Proof of Lemma \ref{lemma1}} \label{proof:lemma1}
We show below that $I'(X; Y) \leq I(X;Y)$. Start by writing
\begin{eqnarray}
& & I(X; Y) - I'(X; Y) \nonumber \\
& & \quad \quad =  \left( H(Y) - H'(Y) \right) - \left( H(V) - H'(V) \right) \nonumber \\
& & \quad \quad = - \int_{-\infty}^{\infty} f_Y(t) \log \left( \dfrac{ f_{Y}(t) }{ f_{Z}(t)} \right ) dt \nonumber \\
& & \quad \quad \quad +  \int_{-\infty}^{\infty} f_V(t) \log \left( \dfrac{ f_{V}(t) }{ f_{G}(t)}\right ) dt \nonumber \\
& &  \quad \quad = - D\Big( f_Y(t) || f_Z(t) \Big) + D\Big( f_V(t) || f_G(t) \Big) \label{eq:diff}
\end{eqnarray}
where $D\Big(p(t) || q(t) \Big)$ is the Kullback-Leibler (K-L)
divergence defined as
\begin{eqnarray}
D\Big( p(t) || q(t) \Big) \triangleq \int_{-\infty}^{\infty} p(t) \log \left( \dfrac{ p(t) }{ q(t)} \right ) dt. \nonumber
\end{eqnarray}
The K-L divergence is always greater than or equal to zero and
convex in pair $\left( p(t) ||q(t) \right)$ \cite{Cover91},
i.e., assuming $p_1(t), p_2(t), q_1(t)$, and $q_2(t)$ are all pdfs,
for $0 \leq \lambda \leq 1$, we have
\begin{eqnarray}
& & D\Big( \lambda p_1(t) + (1 - \lambda) p_2(t) || \lambda q_1(t) + (1 - \lambda) q_2(t) \Big) \nonumber \\
& & \quad \quad \leq \lambda D\Big( p_1(t) || q_1(t) \Big) + (1 - \lambda) D\Big( p_2(t) || q_2(t) \Big). \quad \label{eq:KL_convex}
\end{eqnarray}

For the sake of clarity, we assume that $X$ is from the binary
phase shift keying (BPSK) alphabet, i.e., $X \in \{ \pm \sqrt{P_X}
\}$. Then,
\begin{eqnarray}
f_Y(t) & = & \dfrac{1}{2} \left \{ f_V( t - \sqrt{P_X} ) + f_V( t + \sqrt{P_X} ) \right \} \nonumber \\
f_Z(t) & = & \dfrac{1}{2} \left \{ f_G( t - \sqrt{P_X} ) + f_G( t + \sqrt{P_X} ) \right \}. \nonumber
\end{eqnarray}
Substituting $p_1(t) = f_V( t - \sqrt{P_X} )$, $p_2(t) = f_V( t +
\sqrt{P_X} )$, $q_1(t) = f_G( t - \sqrt{P_X} )$, $q_2(t) = f_G( t
+ \sqrt{P_X} )$ and $\lambda = 0.5$ in (\ref{eq:KL_convex}), we
get
\begin{eqnarray}
D \Big( f_Y(t) || f_Z(t) \Big) & \leq & \dfrac{1}{2} \Big\{ D\Big( f_V( t - \sqrt{P_X} ) || f_G( t - \sqrt{P_X} ) \Big) \nonumber \\
& & +  D\Big( f_V( t + \sqrt{P_X} ) || f_G( t + \sqrt{P_X} ) \Big) \Big\} \nonumber \\
& = & D \Big( f_V(t) || f_G(t) \Big) . \nonumber
\end{eqnarray}
Accordingly, (\ref{eq:diff}) is always greater than or equal to
zero or $I'(X;Y) \leq I(X;Y)$. While this proof is for the binary
alphabet, it is easy to see that the application of the pair-wise
convexity of (\ref{eq:KL_convex}) for any i.u.d. input leads to the same
conclusion.

\section{Derivation of Proposition \ref{proposition1}} \label{appendix:proposition1}
From the pdfs of RVs $V$ and $G$, we can write
\begin{eqnarray}
H'(V) & = & - \int_{-\infty}^{\infty}  f_V(t) \log f_G(t) dt \nonumber \\
& = & \dfrac{1}{2} \log \left( 2 \pi \sigma_{V}^2  \right) + \int_{-\infty}^{\infty} \dfrac{t^2}{2 \sigma_{V}^2 } f_V(t) dt . \label{eq:H'(V)}
\end{eqnarray}
Moreover, we have
\begin{eqnarray}
f_Y(t) & = & \dfrac{1}{2} \left\{ f_V(t -\sqrt{P_X}) + f_V(t + \sqrt{P_X} ) \right\} \nonumber \\
f_Z(t) & = &  \dfrac{1}{2} \left\{ f_G(t -\sqrt{P_X}) + f_G(t + \sqrt{P_X} ) \right\} \nonumber \\
& = & \dfrac{1}{2} \Bigg \{ \dfrac{1} { \sqrt{ 2 \pi  \sigma_{V}^2}  } \exp \left( - \dfrac{ \left(t - \sqrt{P_X} \right)^2 } {  2 \sigma_{V}^2 } \right) \nonumber \\
& & \quad + \dfrac{1}{ \sqrt{ 2 \pi  \sigma_{V}^2}   } \exp \left( - \dfrac{ \left( t + \sqrt{P_X} \right)^2 } {  2 \sigma_{V}^2 } \right) \Bigg \} \nonumber \\
& = & \dfrac{1}{2 \sqrt{2 \pi  \sigma_{V}^2}  } \exp \left( - \dfrac{ (t - \sqrt{P_X} )^2 } {  2\sigma_{V}^2 } \right) \nonumber \\ 
& & \quad \times \left \{ 1 + \exp \left ( \dfrac{ -2 \sqrt{P_X} t} {  \sigma_{V}^2 } \right ) \right \} \nonumber  \\
& = & \dfrac{1}{2 \sqrt{2 \pi  \sigma_{V}^2 } } \exp \left( - \dfrac{ (t + \sqrt{P_X} )^2 } {  2 \sigma_{V}^2 } \right) \nonumber \\
& & \quad \times \left \{ 1 + \exp \left ( \dfrac{ 2 \sqrt{P_X} t} {  \sigma_{V}^2 } \right ) \right \}. \nonumber
\end{eqnarray}
We can write $- \log f_Z(t)$ in two different ways:
\begin{eqnarray}
- \log f_Z(t) & = &  \log 2 + \dfrac{1}{2} \log \left( 2 \pi  \sigma_{V}^2 \right) +  \dfrac{ ( t - \sqrt{P_X} )^2}{ 2 \sigma_{V}^2 } \nonumber \\
& & - \log \left \{ 1 + \exp \left ( \dfrac{ -2 \sqrt{P_X} t} {  \sigma_{V}^2 } \right ) \right \} \nonumber \\
& = &   \log 2 + \dfrac{1}{2} \log \left( 2 \pi  \sigma_{V}^2 \right)  + \dfrac{ ( t + \sqrt{P_X} )^2}{  2 \sigma_{V}^2 } \nonumber \\
& & - \log \left \{ 1 + \exp \left ( \dfrac{ 2 \sqrt{P_X} t} {  \sigma_{V}^2 } \right ) \right \} . \nonumber
\end{eqnarray}
Thus, we have
\begin{eqnarray}
& & - \dfrac{1}{2} \int_{-\infty}^{\infty} f_V(t - \sqrt{P_X}) \log f_Z(t) dt \nonumber \\
& & \quad = \dfrac{1}{2} \left \{ \log 2 + \dfrac{1}{2} \log \left( 2 \pi  \sigma_{V}^2 \right) \right \} \nonumber \\
& & \quad \quad + \dfrac{1}{2}  \int_{-\infty}^{\infty}  \dfrac{ ( t - \sqrt{P_X} )^2}{  2 \sigma_{V}^2 } f_V(t - \sqrt{P_X}) dt \nonumber \\
& & \quad \quad - \dfrac{1}{2} \int_{-\infty}^{\infty} \log \left \{ 1 + \exp \left ( \dfrac{ -2 \sqrt{P_X} t} {  \sigma_{V}^2 } \right ) \right \} f_V(t - \sqrt{P_X}) dt \nonumber \\
& & \quad = \dfrac{1}{2} \left \{ \log 2 + \dfrac{1}{2} \log \left ( 2 \pi  \sigma_{V}^2  \right ) \right \}  +  \dfrac{1}{2}  \int_{-\infty}^{\infty}  \dfrac{ t^2}{  2 \sigma_{V}^2 } f_V(t) dt  \nonumber \\
& & \quad \quad - \dfrac{1}{2} \int_{-\infty}^{\infty} \log \left \{ 1 + \exp \left ( \dfrac{ -2 \sqrt{P_X} t - 2 P_X} {  \sigma_{V}^2 } \right ) \right \} f_{V}(t) dt . \nonumber
\end{eqnarray}
Similarly,
\begin{eqnarray}
& & - \dfrac{1}{2} \int_{-\infty}^{\infty} f_V(t + \sqrt{P_X}) \log f_Z(t) dt \nonumber \\
& & \quad = \dfrac{1}{2} \left \{ \log 2 + \dfrac{1}{2} \log \left ( 2 \pi  \sigma_{V}^2  \right ) \right \} \nonumber \\
& & \quad \quad + \dfrac{1}{2}  \int_{-\infty}^{\infty}  \dfrac{( t + \sqrt{P_X} )^2}{  2 \sigma_{V}^2 } f_V (t + \sqrt{P_X}) dt \nonumber \\
& & \quad \quad - \dfrac{1}{2} \int_{-\infty}^{\infty} \log \left \{ 1 + \exp \left ( \dfrac{ 2 \sqrt{P_X} t} {  \sigma_{V}^2 } \right ) \right \} f_V(t + \sqrt{P_X}) dt \nonumber \\
& & \quad = \dfrac{1}{2} \left \{ \log 2 + \dfrac{1}{2} \log \left ( 2 \pi  \sigma_{V}^2  \right ) \right \} + \dfrac{1}{2}  \int_{-\infty}^{\infty}  \dfrac{ t^2}{  2 \sigma_{V}^2 } f_V (t) dt \nonumber \\
& & \quad \quad - \dfrac{1}{2} \int_{-\infty}^{\infty} \log \left \{ 1 + \exp \left ( \dfrac{ 2 \sqrt{P_X} t - 2 P_X} {  \sigma_{V}^2 } \right ) \right \} f_{V}(t) dt . \nonumber
\end{eqnarray}
Accordingly,
\begin{eqnarray}
& & H'(Y) \nonumber \\
& & \quad = - \int_{-\infty}^{\infty} f_Y(t) \log f_Z(t) dt \nonumber \\
& & \quad = - \dfrac{1}{2} \int_{-\infty}^{\infty} f_V(t - \sqrt{P_X}) \log f_Z(t) dt \nonumber \\
& &  \quad \quad - \dfrac{1}{2} \int_{-\infty}^{\infty} f_V(t + \sqrt{P_X}) \log f_Z(t) dt \nonumber \\
& & \quad = \log 2 + \dfrac{1}{2} \log \left ( 2 \pi  \sigma_{V}^2  \right )  +  \int_{-\infty}^{\infty}  \dfrac{ t^2}{  2 \sigma_{V}^2 } f_V ( t ) dt   \nonumber \\
& & \quad \quad - \int_{-\infty}^{\infty} \dfrac{1}{2} \left [ \log \left \{ 1 + \exp \left ( \dfrac{ -2 \sqrt{P_X} t - 2 P_X} {  \sigma_{V}^2 } \right ) \right \} \right. \nonumber \\ 
& & \quad  \quad \quad \left. +  \log \left \{ 1 + \exp \left ( \dfrac{ 2 \sqrt{P_X} t - 2 P_X} {  \sigma_{V}^2 } \right ) \right \} \right ] f_{V}(t) dt  \nonumber \\
& & \quad = \log 2 + \dfrac{1}{2} \log \left ( 2 \pi  \sigma_{V}^2  \right ) +  \int_{-\infty}^{\infty}  \dfrac{ t^2}{  2 \sigma_{V}^2 } f_V ( t ) dt  \nonumber \\
& & \quad \quad - \int_{-\infty}^{\infty}  \log \left \{ 1 + \exp \left ( \dfrac{ -2 \sqrt{P_X} t - 2 P_X} {  \sigma_{V}^2 } \right ) \right \}f_{V}(t) dt . \nonumber \\ \label{eq:H'(Y)}
\end{eqnarray}
The last equality in (\ref{eq:H'(Y)}) holds because $f_V(t)$ is an
even function. Finally, from (\ref{eq:H'(V)}) and (\ref{eq:H'(Y)}), we arrive at
\begin{eqnarray}
& & I'(X; Y) \nonumber \\
& & \quad = H'(Y) - H'(V)  \nonumber \\
& & \quad = \log 2 \nonumber \\
& & \quad \quad - \int_{-\infty}^{\infty} f_V(t) \log \left \{ 1 + \exp \left ( \dfrac{ -2 \sqrt{P_X} t - 2 P_X} {  \sigma_{V}^2 } \right ) \right \} dt. \nonumber \\ \label{eq:MI-like}
\end{eqnarray}

\sublabon{equation}
Now write $I'(X; Y) = \log 2 - F$ with the new definition
\begin{eqnarray}
F & \triangleq & \int_{-\infty}^{\infty} f_V(t) \log \left \{ 1 + \exp \left ( \dfrac{ -2 \sqrt{P_X} t - 2 P_X} {  \sigma_{V}^2 } \right ) \right \} dt \nonumber \\
& = & 2^{-L} \sum_{i=1}^{2^L} \int_{-\infty}^{\infty} \dfrac{1}{\sqrt{2 \pi \sigma_N^2}} \exp \left( - \frac{(t-m_i)^2}{2\sigma_N^2} \right) \nonumber \\
& & \times  \log \left \{ 1 + \exp \left ( \dfrac{ -2 \sqrt{P_X} t - 2 P_X} {  \sigma_{V}^2 } \right ) \right \} dt \nonumber \\
& = & 2^{-L} \sum_{i=1}^{2^L} \int_{-\infty}^{\infty} \dfrac{e^{ - \tau^2 / 2}}{\sqrt{2 \pi}}  \nonumber \\
& & \times  \log \left \{ 1 + \exp \left ( \dfrac{ -2 \sqrt{P_X} \left( \tau \sigma_N + m_i \right)  - 2 P_X} {  \sigma_{V}^2 } \right ) \right \} d\tau \nonumber \\
& = & 2^{-L} \sum_{i=1}^{2^L} \int_{-\infty}^{\infty} \dfrac{e^{ - \tau^2 / 2}}{\sqrt{2 \pi}}   \log \left \{ 1 + e^{-2 R \rho_i} e^{-2 \phi \sqrt{R} \tau - 2R}  \right \} d\tau \nonumber \\
& = & 2^{-L} \sum_{i=1}^{2^L} \mathrm{E}_{\tau} \left[ \log \left \{ 1 + e^{-2 R \rho_i} e^{-2 \phi \sqrt{R} \tau - 2R}  \right \} \right] \label{eq:pF1}
\end{eqnarray}
where the third equality is obtained with a variable change
$(t-m_i)/\sigma_N = \tau$ and $\rho_i \triangleq m_i /\sqrt{P_X}$, $R \triangleq P_X /
\sigma_V^2$, and $\phi \triangleq \sigma_N / \sigma_V$.
The expression (\ref{eq:pF1}) can also be written as
\begin{eqnarray}
F & = & 2^{-L} \sum_{i=1}^{2^L} \mathrm{E}_{\tau} \left[ \log \left \{ 1 + e^{-2 R \rho_i} e^{-2 \phi \sqrt{R} \tau - 2R}  \right \} \right] \nonumber \\
& = & 2^{- L} \sum_{k = 1}^{2^{L - 1} } \mathrm{E}_{\tau} \Big[ \log \Big\{ {1 + e^{ - 2R\rho_k^+  } e^{ - 2\phi \sqrt R \tau  - 2R} } \Big\} \nonumber \\
& & \quad + \log \Big\{ {1 + e^{2R \rho_k^+  } e^{ - 2\phi \sqrt{R} \tau  - 2R} } \Big\} \Big] \nonumber \\
& = & 2^{ - L} \sum_{k = 1}^{2^{L - 1} } \mathrm{E}_{\tau} \Big[  \log \Big\{ 1 + \left( e^{ - 2R\rho_k^+  }  + e^{ 2R\rho_k^+ } \right) e^{ - 2\phi \sqrt{R} \tau  - 2R} \nonumber \\
& & \quad + e^{ - 4\phi \sqrt{R} \tau  - 4R}  \Big\} \Big] \nonumber \\
& = & 2^{ - (L - 1)} \sum_{k = 1}^{2^{L - 1} } \mathrm{E}_{\tau} \bigg[  \dfrac{1}{2} \log \Big\{ 1 + 2\cosh \big( {2R\rho_k^+  } \big) e^{ - 2\phi \sqrt{R} \tau  - 2R}  \nonumber \\
& & \quad + e^{ - 4\phi \sqrt{R} \tau  - 4R}  \Big\}  \bigg] \label{eq:pF2}
\end{eqnarray} \sublaboff{equation}where $\rho_k^+$'s is the positive-half subset of $\rho_i$'s.

\section{Derivation of the Simple Bounds} \label{simple_bounds}
Due to the convexity of the function, $\mathrm{E}_{\tau} \left[ \frac{1}{2} \log
\left( 1 + 2 \cosh(2R \rho^+) e^{-2 \phi \sqrt{R} \tau} + e^{-4
\phi \sqrt{R} \tau - 4R} \right) \right]$, in $\rho^+$, the upper
bound of $F$ can be found as
\begin{eqnarray}
F & = & 2^{ - (L - 1)} \sum_{k = 1}^{2^{L - 1} } \mathrm{E}_{\tau} \bigg[ \dfrac{1}{2}\log \Big\{ 1 + 2\cosh \big( {2R\rho_k^+  } \big) e^{ - 2\phi \sqrt{R} \tau  - 2R}  \nonumber\\
& & \quad  + e^{ - 4\phi \sqrt{R} \tau  - 4R}  \Big\} \bigg]  \nonumber \\
 & \leq & 2^{ - (L - 1)} \sum_{k = 1}^{2^{L - 1} } \left\{ T \left( |\rho|_{\max}, \theta  \right) \Big\vert_{\theta =  \rho_k^+ } \right\} \nonumber \\
&  = & T \left( |\rho|_{\max}, \theta \right) \Big\vert_{\theta =  {2^{ - (L - 1)} \sum_{k = 1}^{2^{L - 1} } {\rho _k^ +  } } }  \nonumber \\
& = & T \left( |\rho|_{\max}, \theta \right) \Big\vert_{\theta =  |\rho|_{\mathrm{avg}} }  \nonumber \\
& \leq & T \left( |\rho|_{\max}, \theta \right) \Big\vert_{\theta =  \sigma_{\rho} }  \triangleq F^{u1} \label{eq:F_U1}
\end{eqnarray}
where, for a given
$|\rho|_{\max}$, $T(|\rho|_{\max}, \theta)$ represents a straight
line passing through the two points of $\mathrm{E}_{\tau} \Big[
\frac{1}{2} \log \big( 1 + 2 \cosh(2R \theta) e^{-2 \phi \sqrt{R}
\tau} + e^{-4 \phi \sqrt{R} \tau - 4R} \big) \Big]$: at
$\theta=0$ and at $\theta=|\rho|_{\max}$. Also, $\left|\rho  \right|_{\mathrm{avg} } \triangleq 2^{-L}
\sum_{i = 1}^{2^{L} } { | \rho_i |  } = {2^{ - (L - 1)} \sum_{k =
1}^{2^{L - 1} } {\rho _k^ +  } }$. The last inequality in (\ref{eq:F_U1})
is obtained from the Cauchy-Schwarz inequality:
$|\rho|_{\mathrm{avg}} \leq \sigma_{\rho}$.

Another upper bound of $F$ can be also found as
\begin{eqnarray}
F & = & 2^{ - (L - 1)} \sum_{k = 1}^{2^{L - 1} }  \mathrm{E}_{\tau} \bigg[ \dfrac{1}{2} \log \Big\{ 1 + 2\alpha_k e^{ - 2\phi \sqrt{R} \tau  - 2R}  \nonumber \\
& & \quad + e^{ - 4\phi \sqrt{R} \tau  - 4R } \Big\}  \bigg] \nonumber \\
 & \leq &  \mathrm{E}_{\tau} \bigg[ \dfrac{1}{2}\log \bigg\{ 1 + 2\bigg( 2^{ -(L - 1)} \sum_{k = 1}^{2^{L - 1} } \alpha_k  \bigg) e^{ - 2\phi \sqrt{R}\tau  - 2R} \nonumber \\
& & \quad + e^{ - 4\phi \sqrt{R} \tau  - 4R}  \bigg\} \bigg] \nonumber \\
& = & \mathrm{E}_{\tau} \left[ {\dfrac{1}{2} \log \left\{ {1 + 2\alpha_{\mathrm{avg}} e^{ - 2\phi \sqrt{R} \tau  - 2R}  + e^{ - 4\phi \sqrt{R} \tau  - 4R} } \right\}} \right] \nonumber \\ \label{eq:F_U2}
\end{eqnarray}
where $\alpha_k \triangleq \cosh (2R\rho _k^ +  )$ and
$\alpha_{\mathrm{avg}}  \triangleq 2^{ - (L - 1)} \sum_{k =
1}^{2^{L - 1} } {\alpha _k }  = 2^{ - (L - 1)} \sum_{k = 1}^{2^{L
- 1} } {\cosh (2R\rho _k^ +  )} $. The inequality comes from the
concavity of $\mathrm{E}_{\tau} \Big[ \frac{1}{2} \log \Big( 1 +
2 \alpha e^{-2 \phi \sqrt{R} \tau} + e^{-4 \phi \sqrt{R} \tau -
4R} \Big) \Big]$ in $\alpha$. Moreover, since it is an
increasing function of $\alpha$, the last expression of
(\ref{eq:F_U2}) can be further upper-bounded by replacing $\alpha_{\mathrm{avg}}$ with $\alpha' \geq \alpha_{\mathrm{avg}}$ . For example,
note
\begin{eqnarray}
\alpha_{\mathrm{avg}}  & \leq & 2^{-(L - 1)}  \sum_{k = 1}^{2^{L - 1} } { \left( {s\rho _k^ +   + 1} \right)} \nonumber \\
& = & s \left| \rho  \right|_{\mathrm{avg} }  + 1 \leq s \sigma_{\rho}  + 1 \triangleq \alpha' \nonumber
\end{eqnarray}
where $s = \left( \cosh(2R|\rho|_{\max}) - 1 \right) / |\rho|_{\max}$, the slope of a straight line connecting two points of the convex function $\cosh(2R\rho)$, $(0, 1)$ and $(|\rho|_{\max}, \cosh(2R|\rho|_{\max}))$. This gives
\begin{eqnarray}
F & \leq &  \mathrm{E}_{\tau} \bigg[ \frac{1}{2}\log \Big\{ 1 + 2\big( s \sigma_{\rho} + 1 \big) e^{ - 2\phi \sqrt{R} \tau  - 2R} \nonumber \\
& & \quad  + e^{ - 4\phi \sqrt{R} \tau  - 4R}  \Big\} \bigg] \triangleq F^{u2}.
\end{eqnarray}

By using the convexity of the function, $\mathrm{E}_{\tau} \left[ \log \Big( 1
+ e^{-2 R \rho} e^{-2 \phi \sqrt{R} \tau - 2R} \Big) \right]$, in
$\rho$, the lower bound of $F$ is also found as
\begin{eqnarray}
F & = & 2^{ - L} \sum_{i = 1}^{2^L } \mathrm{E}_{\tau} \left[ {\log \left\{ {1 + e^{ - 2R\rho_i } e^{ - 2\phi \sqrt{R}\tau  - 2R} } \right\}} \right] \nonumber \\
  & \geq & \mathrm{E}_{\tau} \bigg[ \log \bigg\{ 1 + \exp \bigg(  \negthickspace \negthinspace - 2R \bigg(  2^{ - L} \sum_{i = 1}^{2^L } \rho_i  \bigg) \bigg) \nonumber \\
  & & \quad \times e^{ - 2\phi \sqrt R \tau  - 2R}  \bigg\}  \bigg] \nonumber \\
& = & \mathrm{E}_{\tau} \left[ {\log \left\{ {1 + e^{ - 2\phi \sqrt{R}\tau  - 2R} } \right\}} \right] \nonumber \\
& = & \mathrm{E}_{\tau} \left[ {\frac{1}{2}\log \left\{ {1 + 2e^{ - 2\phi \sqrt R \tau  - 2R}  + e^{ - 4\phi \sqrt R \tau  - 4R} } \right\}} \right] 
 \triangleq  F^l . \nonumber \\
\end{eqnarray}

\section{Derivation of the Tightened Bounds} \label{tight_bounds}
The tightened bounds are derived in a similar way using the
convexity or concavity of the function except the cluster
identification needs be incorporated. Since $\rho_k = \lambda_n +
\mu_i$, we can rewrite $F$ as
\begin{eqnarray}
  F & = & 2^{ -M} \sum_{n=1}^{2^M } \bigg( 2^{ -(L - M)}     \nonumber \\
  & & \quad \sum_{i = 1}^{2^{L - M} } \mathrm{E}_{\tau} \Big[ \log \Big\{ 1 + e^{ - 2R(\mu_i  + \lambda_n )} e^{ - 2\phi \sqrt{R} \tau  - 2R}  \Big\} \Big]  \bigg)  \nonumber \\
  & = & 2^{ -M} \sum_{n=1}^{2^M }  \bigg( 2^{ - (L - M - 1)}  \nonumber \\
  & & \quad  \sum_{l = 1}^{2^{L - M - 1} }  \negthickspace \negthinspace \mathrm{E}_{\tau} \bigg[ \dfrac{1}{2} \log \Big\{ 1 + 2\cosh \left( {2R\mu_l^+ } \right)  e^{ -2 R \lambda_n }  \nonumber \\
  & & \quad  \quad  \times e^{ - 2\phi \sqrt{R} \tau  - 2R} + e^{ -4 R \lambda_n } e^{ - 4\phi \sqrt{R} \tau  - 4R}  \Big\} \bigg] \bigg)  \nonumber \\
  & \leq & 2^{ -M} \sum_{n=1}^{2^M } \bigg( {2^{ - (L-M-1)} \sum_{l = 1}^{2^{L-M-1} } \negthickspace \bigg\{ T_n (|\mu|_{\max}, \theta  ) \Big\vert_{\theta =  \mu_l^+ } \bigg\} } \bigg)   \nonumber \\
  & = & 2^{ -M} \sum_{n=1}^{2^M } \left\{ T_n \left(|\mu|_{\max}, \theta \right) \Big\vert_{\theta =  2^{ - (L - M - 1)} \sum_{l = 1}^{2^{L - M - 1} }\negthickspace \negthinspace {\mu_l^+ }  }  \right\} \nonumber \\
  & = & 2^{ -M} \sum_{n = 1}^{2^M } \left\{ T_n (|\mu|_{\max}, \theta ) \Big\vert_{\theta =  |\mu|_{\mathrm{avg}} } \right\} \nonumber \\
  & \leq & 2^{ -M} \sum_{n = 1}^{2^M } \left\{ T_n (|\mu|_{\max}, \theta ) \Big\vert_{\theta =  \sigma_{\mu} } \right\}  \triangleq F_M^{u1}
\end{eqnarray}
where $\mu_l^+$'s form the positive-half subset of $\mu_i$'s and,
for a given $|\mu|_{\max}$, $T_n(|\mu|_{\max}, \theta)$ is a
straight line that passes through the convex function $\mathrm{E}_{\tau} \Big[
\frac{1}{2} \log \Big\{ 1 + 2\cosh \left( 2R \theta  \right)
e^{ -2 R \lambda_n } e^{ - 2\phi \sqrt{R} \tau  - 2R}  + e^{ -4 R
\lambda_n } e^{ - 4\phi \sqrt{R} \tau  - 4R}  \Big\} \Big]$ at $\theta = 0$ and
$\theta=|\mu|_{\max}$. Moreover, $|\mu|_{\mathrm{avg}} \triangleq
2^{ - (L - M)} \sum_{i = 1}^{2^{L - M} } | \mu_i | = 2^{ - (L - M
- 1)} \sum_{l = 1}^{2^{L - M - 1} } \negthinspace \mu_l^+ $. The last inequality
also follows from $|\mu|_{\mathrm{avg} } \leq \sigma_{\mu}$, and
note $\sigma_{\mu} = \sqrt{\sigma_{\rho}^2 - \sigma_{\lambda}^2}$
and $|\mu|_{\max} = |\rho|_{\max} - |\lambda|_{\max}$.

Another form of tightened upper bound of $F$ is obtained as
\begin{eqnarray}
F & = & 2^{ -M} \sum_{n=1}^{2^M } \bigg( 2^{ -(L-M-1)}  \nonumber \\
& & \quad  \sum_{l=1}^{2^{L-M-1} } \negthickspace \negthinspace \mathrm{E}_{\tau} \bigg[ \dfrac{1}{2} \log \Big\{ 1 + 2 \cosh \left( {2R\mu_l^+  } \right) e^{ - 2R\lambda_n }  \nonumber \\
& & \quad \quad \times e^{ - 2\phi \sqrt{R}\tau  - 2R} + e^{ -4R \lambda_n } e^{ -4\phi \sqrt{R} \tau  - 4R}  \Big\} \bigg]  \bigg)  \nonumber \\
  & = & 2^{ -M} \sum_{n=1}^{2^M } \bigg( 2^{ -(L-M-1)}  \nonumber \\
& & \quad  \sum_{l=1}^{2^{L-M-1} } \negthickspace \negthinspace \mathrm{E}_{\tau} \bigg[ \dfrac{1}{2} \log \Big\{ 1 + 2\beta_l e^{ - 2R \lambda_n } e^{ - 2\phi \sqrt{R} \tau  - 2R} \nonumber \\
& & \quad \quad  + e^{ -4R \lambda_n } e^{ -4\phi \sqrt{R} \tau  - 4R}  \Big\} \bigg]  \bigg)  \nonumber \\
  & \leq & 2^{ -M} \sum_{n=1}^{2^M } \mathrm{E}_{\tau} \bigg[ \dfrac{1}{2} \log \bigg\{ 1 + 2\bigg( 2^{ -(L-M-1)} \sum_{l=1}^{2^{L-M-1} } \negthickspace \beta_l   \bigg) \nonumber \\
  & & \quad \times  e^{ -2R\lambda_n } e^{ - 2\phi \sqrt{R}\tau  - 2R}  + e^{ -4R \lambda_n } e^{ -4\phi \sqrt{R}\tau  - 4R}  \bigg\} \bigg]   \nonumber \\
  & = &  2^{ -M} \sum_{n=1}^{2^M } \mathrm{E}_{\tau} \bigg[ \dfrac{1}{2} \log \Big\{ 1 + 2\beta_{\textrm{avg}} e^{ - 2R\lambda_n } e^{ - 2\phi \sqrt{R} \tau  - 2R}  \nonumber \\
  & & \quad + e^{ - 4R\lambda_n } e^{ - 4\phi \sqrt{R} \tau  - 4R}  \Big\} \bigg]  \nonumber \\
  & \leq & 2^{ -M} \sum_{n=1}^{2^M } \mathrm{E}_{\tau} \bigg[ \dfrac{1}{2} \log \Big\{ 1 + 2\beta' e^{ - 2R\lambda_n } e^{ - 2\phi \sqrt{R} \tau  - 2R}  \nonumber \\
  & & \quad + e^{ - 4R\lambda_n } e^{ - 4\phi \sqrt{R} \tau  - 4R}  \Big\} \bigg]  \nonumber \\
  & = & 2^{ -M} \sum_{n=1}^{2^M } \mathrm{E}_{\tau} \bigg[ \dfrac{1}{2} \log \Big\{ 1 + 2\big( {s_M \sigma_{\mu}  + 1} \big) e^{ - 2R\lambda_n }  \nonumber \\
  & & \quad \times e^{ - 2\phi \sqrt{R} \tau  - 2R} + e^{ - 4R\lambda_n } e^{ - 4\phi \sqrt{R}\tau  - 4R}  \Big\} \bigg]  \triangleq F_M^{u2} \nonumber \\
\end{eqnarray}
where $\beta_l \triangleq \cosh \left( {2R\mu_l^+ } \right)$,
$\beta_{\textrm{avg}}  \triangleq 2^{ - (L - M - 1)} \sum_{l =
1}^{2^{L - M - 1} } \negthinspace {\beta_l} = 2^{ - (L - M - 1)} \sum_{l =
1}^{2^{L - M - 1} } \negthickspace {\cosh (2R\mu_l^+  )}$ and
\begin{eqnarray}
\beta_{\textrm{avg}}  & \leq & 2^{-(L - M - 1)} \sum_{k = 1}^{2^{L - M - 1} } \negthickspace {\left( {s_M \mu_k^+ + 1} \right)} \nonumber \\
& = & s_M \left| \mu  \right|_{\mathrm{avg}}  + 1 \leq s_M \sigma_{\mu}  + 1  \triangleq \beta' \nonumber
\end{eqnarray}
which is based on a straight line connecting two points of the
convex function $\cosh(2R\mu)$, $(0,1)$ and $(|\mu|_{\max},
\cosh(2R|\mu|_{\max})$, having a slope $s_M =
\left(\cosh(2R|\mu|_{\max})-1 \right) / |\mu|_{\max}$.

The tightened lower bound of $F$ based on cluster identification
is obtained as
\begin{eqnarray}
  F & = & 2^{ -M} \sum_{n = 1}^{2^M } \bigg( 2^{ -(L-M)} \nonumber \\
  & & \quad  \sum_{i=1}^{2^{L-M} } \mathrm{E}_{\tau} \Big[ \log \Big\{ 1 + e^{ - 2R( \mu_i + \lambda_n  )} e^{ - 2\phi \sqrt{R} \tau  - 2R}  \Big\} \Big]  \bigg)  \nonumber \\
  & = & 2^{ -M} \sum_{n=1}^{2^M } \bigg( 2^{ -(L-M)}  \nonumber \\
  & & \quad \sum_{i = 1}^{2^{L-M} } \mathrm{E}_{\tau} \Big[ \log \Big\{ 1 + e^{ - 2R\mu_i } e^{ - 2R\lambda_n } e^{ - 2\phi \sqrt{R}\tau  - 2R}  \Big\} \Big]  \bigg)  \nonumber \\
  & \geq & 2^{ -M} \sum_{n=1}^{2^M } \mathrm{E}_{\tau} \bigg[ \log \bigg\{ 1 + \exp \bigg( \negthickspace \negthinspace - 2R \bigg( 2^{ -(L-M)} \negthinspace \sum_{i = 1}^{2^{L - M} } \negthickspace \mu_i  \bigg)  \bigg) \nonumber \\
  & & \quad \times e^{ - 2R\lambda_n } e^{ - 2\phi \sqrt{R}\tau  - 2R}  \bigg\} \bigg]  \nonumber \\
  & = & 2^{ -M} \sum_{n=1}^{2^M } { \mathrm{E}_{\tau} \left[ {\log \left\{ {1 + e^{ - 2R\lambda_n } e^{ - 2\phi \sqrt{R}\tau  - 2R} } \right\}} \right]}  \nonumber \\
  & = & 2^{ -(M-1)} \sum_{k = 1}^{2^{M - 1} } \mathrm{E}_{\tau} \bigg[ \dfrac{1}{2} \log \Big\{ 1 + 2\cosh \big( 2R\lambda_k^+  \big) \nonumber \\
  & & \quad \times e^{ - 2\phi \sqrt{R} \tau  - 2R}  + e^{ - 4\phi \sqrt{R} \tau  - 4R}  \Big\} \bigg]  \triangleq F_M^l
\end{eqnarray}
where $\lambda_k^+$'s form the positive-half subset of $\lambda_n$'s.

\biographynophoto
{Seongwook Jeong}
received the B.S. degree with honor in electrical engineering from Seoul National University, Seoul, Korea, in 2004 and the M.S. degree from the University of Southern California, Los Angeles in 2005. He is currently working toward the Ph.D. degree at the University of Minnesota, Minneapolis. Since 2006, he has been a member of the Communications and Data Storage Laboratory, University of Minnesota. He also worked at Link-A-Media Devices Corp. and LSI Logic Corp. during the summer of 2008 and 2011, respectively. His research interests include information theory, coding theory, and equalization/detection with application to wireless communication and data storage systems.

\biographynophoto
{Jaekyun Moon}
is a Professor of Electrical Engineering at KAIST. Prof. Moon received the B.S.E.E. degree with high honor from Stony Brook University and the M.S. and Ph.D. degrees in Electrical and Computer Engineering at Carnegie Mellon University. From 1990 through early 2009, he was with the faculty of the Department of Electrical and Computer Engineering at the University of Minnesota, Twin Cities. Prof. Moon's research interests are in the area of channel characterization, signal processing and coding for data storage and digital communication. He received the 1994-1996 McKnight Land-Grant Professorship from the University of Minnesota. He also received the IBM Faculty Development Awards as well as the IBM Partnership Awards. He was awarded the National Storage Industry Consortium (NSIC) Technical Achievement Award for the invention of the maximum transition run (MTR) code, a widely-used error-control/modulation code in commercial storage systems. He served as Program Chair for the 1997 IEEE The Magnetic Recording Conference (TMRC). He is also Past Chair of the Signal Processing for Storage Technical Committee of the IEEE Communications Society. In 2001, he cofounded Bermai, Inc., a fabless semiconductor start-up, and served as founding President and CTO. He served as a guest Editor for the 2001 IEEE J-SAC issue on Signal Processing for High Density Recording. He also served as an Editor for IEEE Transactions on Magnetics in the area of signal processing and coding for 2001-2006. He worked as consulting Chief Scientist at DSPG, Inc. from 2004 to 2007. He also worked as Chief Technology Officer at Link-A-Media Devices Corp. in 2008. He is an IEEE Fellow.

\end{document}